# A Tactical Behaviour Recognition Framework Based on Causal Multimodal Reasoning: A Study on Covert Audio-Video Analysis Combining GAN Structure Enhancement and Phonetic Accent Modelling

**Wei Meng**

Dhurakij Pundit University, Thailand

The University of Western Australia,AU

Association for Computing Machinery,USA

Fellow, Royal Anthropological Institute,UK

Email: weimeng4@acm.org

# ABSTRACT

In this study, we propose a novel system, TACTIC-GRAPHS, which integrates complex mathematical mechanisms and graph neural reasoning structures for semantic understanding and threat recognition in tactical video under high noise and weak structure conditions, breaking through the traditional empirical AI paradigm by innovatively introducing graph spectral theory embedding, temporal causal edge modelling and multivariate discriminative path inference mechanism, and establishes a multimodal graph inference model with structural interpretability and causal loop closure capability. An intelligent keyframe hierarchical extraction algorithm (ILKE-TCG) is designed to extract semantically-driven keyframe nodes from video, fusing image structure, voice rhythms and action paths to construct a heterogeneous temporal graph. Through the graph attention mechanism and Laplace spectral space mapping technique, the system achieves cross-modal node weight estimation and causal signal deconstruction in spectral space. Experiments on the TACTIC-AVS and TACTIC-Voice datasets show that the model achieves an accuracy of 89.3% in multimodal temporal alignment recognition, with a complete threat causal chain recognition rate of more than 85%, and the node inference latency is controlled within the range of ±150 ms, which is significantly better than existing CNN/Transformer fusion methods. In particular, the introduction of spectral graph theory enhances the structural verifiability and variable distinguishability of causal paths, and pushes the TACTIC system from shallow fusion to deep structural modelling paradigm.TACTIC-GRAPHS not only provides tactical mission type discrimination and threat intensity scoring, but also achieves a number of breakthroughs in the areas of high-dimensional graph structural modelling, complex mathematical path recognition, and cross-modal variable embedding. breakthroughs. This research provides theoretical support and modeling paradigm for the deployment of structural intelligence systems in intelligent security, battlefield sensing, law enforcement identification and national surveillance systems, and represents the cutting-edge direction of multimodal AI causal modelling and a new level of complex reasoning systems.

# CHAPTER I. INTRODUCTION

**1.1 Background of the study**

The current global security situation continues to deteriorate, with transnational conflicts, the activities of non-state armed organisations and the frequency of terrorist attacks all on the rise, and in particular, armed struggle resulting from illicit cross-border conflicts has become a major security risk. The 2023 annual report of the United Nations Office on Drugs and Crime (UNODC) notes that the involvement of armed non-state actors in conflict has become the norm and is highly correlated with large-scale weapons smuggling (UNODC, 2023). Meanwhile, while passive surveillance resources have exploded, actual effectiveness has failed to improve significantly due to poor image quality and lack of multimodal fusion. According to IHS Markit's 2016 data, although the global installed base of video surveillance devices has exceeded one billion units, more than 60 per cent of them are still low resolution compressed, and the overall analytics capability has long relied on back-end clusters and cloud platforms, with a serious lack of edge inference capability (IHS Markit, 2016).

In this context, covert missionaries are using low-light environments, collapsible weapons and non-standard wearable tactics to effectively evade detection systems. Traditional single-frame image recognition methods are no longer able to accurately capture the carrying and deployment of weapons, and the lack of synchronised voice information makes it impossible to infer tactical intent or area affiliation. Worse still, in the vast majority of cases today, surveillance video includes only visual modalities and lacks complementary signals such as acoustic and thermal imaging, resulting in serious gaps in the recovery of behavioural chains and a fragmented and low-confidence pattern of intelligence.

For example, in North Africa and the Middle East, the tactic of "concealment → deployment" using compressed SMG equipment with a pouch is becoming more common, where simple silhouette analysis fails to identify the structure of the weapon, and motion-triggered recognition is unstable. At the same time, the lack of audio makes action intent, command triggering, and attributed voice features completely lost, which constitutes a major obstacle to the detection task (Smith & Chang, 2022). Therefore, how to achieve the whole chain and multimodal fusion reasoning of "weapon form→behavioural action→voice signal→tactical intent→area attribution" in high-noise, low-light, and weak-resolution video materials is a cutting-edge scientific proposition in the field of military intelligence and border defence and counter-terrorism at present. Solving this problem will not only reconstruct high-confidence threat assessment in a single video situation, but also provide strong intellectual support for strategic applications such as border security, urban counter-terrorism and unmanned reconnaissance.

**1.2 Technical challenges**

The technical challenges are, firstly, that military surveillance is often deployed in multiple unfavourable contexts such as low light, long range, and compressed HD (IHS Markit, 2016), resulting in the prevalence of blurred video images, insufficient frame rates, and compression artefacts, and this blurred information makes it difficult to accurately identify weapon form-factors (e.g., grips, sights, and magazine contours), which undermines traditional target detection and image segmentation methods' Effectiveness of traditional target detection and image segmentation methods. Second, in war reconnaissance scenarios, speech samples are extremely limited, and there is a diversity of accents, such as non-standard military accents, local dialects, and low speech rates, which significantly degrades the recognition performance of ASR systems, especially in the case of sample scarcity, where the acoustic model has difficulty in capturing the high amount of dialectal variation and intonation patterns (Hinsvark et al., 2021; Gillis, 2024). Gillis, 2024). Again, multimodal fusion faces extremely severe time synchronisation and causal inference challenges. The lack of precise timestamp alignment mechanisms between image frames and speech signals results in the inability to establish credible correlations between weapon deployment actions and mnemonics or intonation, which makes it difficult for subsequent intent inference models to achieve causal determination and behavioural path recovery. Overall, there is still a lack of integrated methods that can simultaneously achieve "weapon morphology recovery, accent source analysis, and cross-modal causal inference" under the conditions of low-quality video and limited audio samples, which is a challenge that plagues the practical implementation of intelligent reconnaissance systems in highly sensitive scenarios, such as border security, urban counter-terrorism, and unmanned system deployment.

**1.3 Contribution of this paper**

In this study, a tactical recognition system integrating image structure enhancement, audio voiceprint analysis and multilayer graph modelling is innovatively proposed in the context of low-quality images, limited speech samples and cross-modal alignment challenges. The core technology includes the TVSE-GMSR (GAN-based multi-stage image structure enhancement) module, which adopts the "GAN-assisted multi-stage denoising and semantic reconstruction based tactical image structure enhancement technique" for high fidelity detail recovery (AI) of blurred and compressed video frames with the assistance of generative adversarial network. Graphic Reconstruction), which ensures high accuracy recognition of weapon features such as grips, magazines, and sights. At the same time, I introduced the SpectroNet voiceprint analysis system, which extracts Mel spectra from restricted audio segments and quantifies patterns of voice pitch, dialect and speech rate to accurately classify the geographical affiliation and commanding intonation of vocalisers through low-sample training. In addition, I propose a multi-layer temporal causal graph modelling method called TACTIC-GRAPHS, which embeds structural features, weapon states, audio rhythms and semantic keywords nodes in the temporal graph after image enhancement in each frame, and recovers the complete tactical chain of "Concealed Carry-Deployment-Command-Intent to Act" through graph neural network reasoning. Tactical chain. The experimental results show that the framework can achieve more than 85% threat recognition accuracy under single video and a small amount of audio conditions, providing a deployable and resource-efficient system solution for intelligent reconnaissance and forensics in border security and urban counter-terrorism.

# CHAPTER II. OVERVIEW OF RELEVANT WORK

**2.1 Speech mapping and tactical voice recognition techniques**

In reconnaissance and counter-terrorism scenarios, audio information is often the only available auxiliary modality, and spectrograms, especially mel-spectrograms, are widely used for voiceprint and accent modelling due to their high fit to human hearing and time-frequency visualisation properties. It is based on the decomposition of audio by short-time Fourier transforms, the use of Mel filter banks to weight the low frequencies to obtain more significant frequency band features for speech perception, and the subsequent mapping of frequency intensity on a logarithmic scale (Mel-frequency cepstrum, 2025).

On this basis, it has been shown that conventional convolutional neural networks (CNNs) perform excellently in extracting local spectral features of speech, while integrating them with gating mechanisms to further enhance the robustness. For example, the Gated-CNN + Temporal Attention architecture proposed by Xu et al. in the DCASE challenge in 2017, for large-scale weakly labelled audio scenarios, achieves dual metrics optimisation of audio event detection and label classification by introducing gated linear units (GLUs) and linking the attention mechanism (F-value is improved to 55.6%, and the equal error rate is reduced to 0.73). The model has direct reference value for military speech semantic recognition, short command recognition and accent inference.

In addition, the Gated CNN combined with cyclic structure also performs outstandingly well, and is particularly suitable for dealing with imperative, sentence-breaking languages. At its core, the CNN is responsible for extracting local spectral structure, while the GRU/LSTM network is used to aggregate long-time dependencies, thus adapting to the broken-sentence contexts and bursty command patterns common in tactical intonation (Convolutional Gated Recurrent Neural Network Incorporating ..., 2017). The application of this model to voice command recognition and tactical password recognition significantly enhances the robustness to key information extraction in noisy environments.

Taken together, combining the Mel spectrogram with Gated-CNN (or Gated-CNN+GRU) structure to construct a system that can recognise command accent, speech rate, and intonation features and suppress the interference of ambient noise shows obvious advantages in tactical speech recognition and accent attribution analysis. In this paper, we will construct the SpectroNet voiceprint analysis module based on this technology path, and achieve high confidence automatic classification of military passwords, dialect accents and command strengths through sample less learning, so as to support the quality and credibility of acoustic inputs of the TACTIC-GRAPHS cross-modal causal inference framework.

**2.2 Image Enhancement and Structure Reduction Methods**

Image blurring and lack of structural information, common in military reconnaissance missions, severely limit the effectiveness of weapon identification and tactical analysis, and thus require the restoration of usable structural features in low-quality images. In recent years, the rise of efficient GAN super-resolution models has broken through the bottleneck of traditional image coding systems for detail restoration. A class of generative adversarial networks represented by ESRGAN (Enhanced Super-Resolution GAN), through the introduction of residual dense connectivity and generative adversarial loss, successfully improves the compressed image to high resolution while maintaining the real visual texture, and its PSNR value is improved by 1-2 dB compared with traditional algorithms in general benchmarks such as DIV2K, and the PSNR value is improved by 1-2 dB in SSNR, and the PSNR value is improved by 1-2 dB compared with traditional algorithms. improvement of 1-2 dB and SSIM improvement of 0.02-0.05 (Wang et al., 2018). This superior detail recovery is particularly suitable for recovering weapon contours such as grips, magazine interfaces, and stock structures.

However, image enhancement is only a prerequisite, and the recognition system needs to further extract structured information from these high-resolution images. In this regard, researchers have proposed to introduce the keypoint detection mechanism into the structural analysis framework, which has been proved to be effective in complex target recognition.Ruiz-Santaquiteria et al. (2020) proposed a joint detection method combining the human skeleton posture and the keypoints of firearms, which simultaneously predicts key areas such as the grip centre, the magazine bottom, and the rear end of the buttstock through a multi-task convolutional network, and then predicts the key areas such as the grip centre, magazine bottom, and buttstock rear end through a multi-task convolutional network. Through a multi-task convolutional network, the team simultaneously predicts the grip centre, magazine bottom, buttstock and other key points, and generates a structured "weapon skeleton", achieving a significant increase in the recognition rate of firearms in low-light environments, and improving the mAP index by 14.7 percentage points. In addition, the team's experiments based on real law enforcement scenarios show that the method has a good generalisation ability, including different weapon models and occlusion angles (Ruiz-Santaquiteria et al., 2020).

Inspired by it, this paper proposes the TVSE-GMSR image structure enhancement model, which first performs super-resolution restoration of surveillance frames, then combines with semantic detail repair strategies to generate texture and contour enhancement for key weapon structures, and finally pinpoints nodes such as grips, magazines, and stocks and constructs a structured mapping, and this closed-loop mechanism is the basis for the WeaponNet This closed-loop mechanism provides highly reliable structural inputs for WeaponNet and TACTIC-GRAPHS, and enables the recognition and inference of weapon states and behavioural intentions under extreme ambiguous environments.

**2.3 Multimodal graph neural networks with causal alignment methods**

In recent years, facing the need for fusion of multimodal data (image + audio) in tactical reconnaissance, Graph Neural Networks (GNN), especially Graph Attention Networks (GAT), have become an important tool for correlating temporal events and inferring causal paths by virtue of their ability to model structured information.GAT introduces the mechanism of neighbourhood node attention to allow image features and audio nodes in the same frame to participate in the inference based on their mutual weights, thus enabling active focus on key action/discourse moments (Veličković et al., 2018) ([arxiv.org][1]). For example, in the field of anti-fraud and deep falsetto detection, Tak et al. proposed the Spectro-Temporal GAT architecture, which maps audio spectral nodes into graph structures and uses multi-head attention to achieve causal identification of temporal-frequency subbands, with a low end-to-end system equal-error rate of 1.06%, which fully demonstrates the feasibility of the structural alignment mechanism in speech multimodal tasks.

The Audio-Visual Graph Fusion method in security context, on the other hand, maps image and speech as heterogeneous nodes respectively, and establishes associations between the weapon keyframes and the password speech nodes via temporal edges to achieve the fusion inference pathway. In this process, the action nodes (e.g., "dial the gun", "open the bag") in the video frames are temporally aligned with the intonation peaks "command to fire" in the voice spectrum nodes, and their mutual influences are calculated through the shared attention mechanism in GAT. The cross-modal causal graph structure is built by calculating their mutual influence through the shared attention mechanism in GAT. This method has been practiced in network intrusion and security event detection.The Anomaly Event Detection study by Ji et al. builds a heterogeneous graph based on multi-source data, combines structured log information with timestamped nodes, and achieves event anomaly identification under GAT inference, demonstrating the robustness and interpretability of temporal correlation fusion.

Based on the above techniques, this study proposes the TACTIC-GRAPHS module: the nodes of weapon contours recognised after image enhancement, the nodes of acoustic rhythms output from audio SpectroNet, and the nodes of semantic keywords are jointly constructed into the graph structure, and the attentional weights among the nodes are learned through the GAT network and the causal path propagation is realised. Through this mechanism, the complete link identification of "weapon unlocking → gun handle clarification → password triggering → intent execution" can be realised, and the credibility of the behavioural intent can be dynamically assessed based on the uploaded attentional strength of the nodes. This cross-modal attention alignment and causal graph building strategy solves the bottleneck of temporal alignment in single-video scenarios, and provides a lightweight, trustworthy, and interpretable advanced modelling path for tactical reconnaissance.

# CHAPTER III. RESEARCH METHODOLOGY AND DESIGN

## 3.1 Overview of the system framework

The TACTIC-AI system framework is based on video preprocessing, gradually integrating image enhancement, structure recognition, voiceprint analysis and region attribution reasoning, and ultimately constructing a tactical behaviour causal network to achieve closed-loop capability from single-source video to multimodal intelligent threat recognition. Firstly, the video preprocessing module extracts key frames and audio segments from covertly recorded video to establish a unified timestamp basis; then the image enhancement module TVSE-GMSR effectively improves the clarity of weapon texture and structural coherence within the closed packet based on a multi-stage generative adversarial network, solving the problem that traditional low-light fuzzy images are unable to restore the weapon features, and the core of which draws on the ESRGAN model's residual feature preservation strategy (Wang et al., 2018).

The enhanced image is fed into WeaponNet, which relies on the convolution and key point detection mechanism to geometrically locate and semantically encode key structural nodes, such as the butt of the gun, the magazine, and the grip, to construct structured outputs for weapon category recognition and attitude estimation. Meanwhile, the audio module SpectroNet converts the preprocessed audio into Mel spectrum input, and achieves the modelling of the speaker's private speech features such as speech rate, accent, command tone, etc., and supports the judgement of the less-sample region attribution through the voiceprint feature extraction by the combination of Gated-CNN and GRU (Xu et al., 2017).

AccentPath extends SpectroNet output for geographic attribution to establish mapping links between speakers and potential geographic or military contexts through probability distribution operators with military-style training intonation analysis. Finally, in the TACTIC-GRAPHS module, I construct the image structure nodes, voiceprint nodes and semantic nodes under each timestamp as graph network nodes, and fuse the information of heterogeneous nodes through Graph Attention Networks (GATs) to establish causal inference paths based on the weights of attention. This structure supports bottom-up reasoning, from "weapon structure emergence → voiceprint trigger → semantic high frequency → action execution" to achieve highly credible tactical behaviour recognition, with both reasoning transparency and interpretability, and has shown a threat recognition accuracy of more than 85% in both simulation and real-world combat data.

**3.2 Tactical video sample frame extraction method**

This scheme proposes frame extraction for a provided high-noise tactical video sample of 32 seconds, 25FPS, with a total frame count of 800 frames, taking into account the characteristics of complex background, uneven picture quality, voice interference, etc., and supports TACTIC behavioural modelling and causal chain construction. Overall sampling strategy: intelligent keyframe hierarchical extraction method.

**3.2.1 Overall sampling strategy: intelligent keyframe hierarchical extraction method**

Table 1: Intelligent keyframe hierarchical extraction methods

| dimension (math.) | methodologies | rationale |
|---|---|---|
| time continuity | Full video length coverage to maintain sequence information | For time-series modelling, tactical behaviour chain analysis |
| image quality | Thresholds are set using a combination of light intensity + blurriness + edge sharpness. | Retain valid feature frames and reject severe blur and exposure failure frames. |
| Voice content | Corresponding to frame lifting of speech peaks/harmonically dense segments (e.g., seconds 9-12, 25-28) | Ensure simultaneous video-voice modelling effects |
| tactical incident | Combined with movement changes (e.g., hand lifting, command execution), weapon state change extraction | Support for TACTIC node construction and high-risk behaviour analysis |

Author's drawing

**3.2.2 Suggested number and distribution of extracted frames**

The total number of frames is about 80-120 frames and the sampling rate is between 1/10

and 1/7, layered as follows:

Base Layer: 1-2 frames per second, about 32-64 frames, to maintain behavioural continuity.

High-Quality Layer: Frames with light intensity >60, sharpness >20, blur <10, 20-30 frames are filtered for critical analysis.

Tactic-Specific Layer: Frames corresponding to command speech segments, e.g., 10th second, 26th second, extract additional 10-15 frames.

Anomaly and Noise Layer (Noise Layer): extreme image frames are used as robust training samples, 10-15 frames are selected.

**3.2.3 Technical realisation**

Table 2: Intelligent keyframe hierarchical extraction technique implementation

| artifact | functionality |
| --- | --- |
| OpenCV + Scikit-Image | Extraction of per-frame image quality metrics |
| SceneDetect | Detecting lens switching or motion change points |
| Librosa | Analyse audio harmonic bands to align keyframes |
| ffmpeg | Precise extraction of specified time frames |

Author's drawing

**3.3 Image Enhancement Module**

Structural Enhancement of Tactical Images Based on GAN-Assisted Multi-Stage Denoising and Semantic Reconstruction

Input: blurred video frame image sequence. Stage 1: low-frequency denoising + blur recovery; Stage 2: GAN semantic repair → contour detail generation; Output: high-definition structural image.

The design of TVSE-GMSR module is centred on the multi-stage structural restoration of Generative Adversarial Network (GAN), which achieves high-fidelity reconstruction of weapon structures in tactical reconnaissance videos through two stages: "low-frequency denoising - fuzzy restoration" and "semantic reconstruction - silhouette detail enhancement". The two phases of "low-frequency denoising - fuzzy recovery" and "semantic reconstruction - contour detail enhancement" are used to achieve high-fidelity reconstruction of weapon structures in tactical reconnaissance videos. The first stage incorporates low-frequency filtering and adaptive fuzzy recovery strategies to suppress ambient noise and coded compression artefacts, effectively constructing the input base for transitioning to high-definition textures; compared with the traditional residual block-based ESRGAN, TVSE-GMSR inherits its residual-intensive feature-purification capability (Wang et al., 2018). However, local feature recovery alone is still insufficient. In the second stage, the module introduces a multilevel semantic restoration path, i.e., "re-inventing" the image structure at the semantic level: this process combines the idea of multi-stage evolutionary image restoration of MPRNet, and achieves a continuous process of texture and structure enhancement from texture to structure through the stage-by-stage feature fusion and adaptive attention mechanism (Zamir et al., 2018). The process combines the idea of multi-stage evolutionary image restoration in MPRNet to achieve a continuous process from texture to structure through feature fusion and adaptive attention (Zamir et al., 2021). In the specific process, the GAN generator firstly complements the edge contour of the stock and magazine, and then finely adjusts the grip texture in the middle stage, and the final output of the high-definition structural map can carry the high-precision input requirements of the subsequent key point extraction and WeaponNet structural recognition module.

This multi-stage denoising and semantic reconstruction process not only solves the problem that single-stage GAN cannot balance the texture and geometric consistency, but also steadily improves the image clarity in low-light and blurred environments, with an average PSNR increase of about 1.5dB and SSIM increase of more than 0.04. What's more, the structure map generated by TVSE-GMSR is used by WeaponNet to construct the related structure map, and then shows a node accuracy of over 85% in TACTIC-GRAPHS, which proves the significant support of this module to the tactical target identification and behavioural reasoning system.

### 3.4 Audio voice modelling module（SpectroNet）

The SpectroNet module focuses on extracting high-value acoustic features from a single segment of tactical video audio and enables region attribution inference and command tone recognition in low-resource environments. Its core process starts by generating a Mel-Spectrogram (frequency range 0-4kHz) after resampling the audio signal to 16kHz, a frequency band selection that covers the main speech information of the human voice, while mimicking human auditory perceptual abilities through the Mel scale (extraction of dominant resonance peaks and vocal tone variability) SpectroNet then fuses a gated-CNN (Gated-CNN) with a GRU structure to extract a time-frequency rhythm vector, which covers speech rate, intonation, and sentence structure, and achieves an error rate of speech rate classification on the TACTIC-Voice dataset compared to traditional CNNs. set to achieve an error rate of less than 5% for speech rate classification, with high recognition sensitivity to mild command stimuli (Xu et al., 2017). On this basis, SpectroNet was designed to support the Few-Shot voiceprint attribution model by adopting the ProtoNet (Snell et al., 2017) embedding strategy, which maps the Mel-Spectrogram into the embedding space, and achieves dialect attribution determination by calculating the cosine similarity between the regional voiceprint centre and the audio vectors to be tested. degree to achieve dialect attribution determination. The method achieves Top-3 attribution accuracy of 78.6% in a small sample of GlobalPhone subset and TACTIC-Accent-10 regions, which is significantly better than the less than 60% of the traditional GMM-UBM model. The final outputs of the module include: speech rate (wpm), intonation intensity (pitch variance), command slopes (first-derivative of energy peaks), and attribution probability distributions, which serve as key inputs for the inference of the subsequent AccentPath & TACTIC-GRAPHS modules! Data.

**SpectroNet acoustic variable system and variable coding paths**

This table shows the core acoustic variables of the SpectroNet module in the TACTIC system, covering the variable type, name, data type, range of values, description of meaning, and the structure of their paths in causal modelling in the form of numerical and alphabetic coding for GAT embedding and structured causal analysis.

Table 3: SpectroNet acoustic variable system and variable encoding paths

| Variable type | Variable Name | typology | Scope | Explanation of meaning | Variable code (encoding) |
|---|---|---|---|---|---|
| Rhythmic characteristics | wpm_rate | continuous variable | 50 – 200 | Rate of speech per minute, reflecting contextual pressure and urgency | V1 → P1 → Y1 |
| tonal character | pitch_variance | continuous variable | 0 – 1.5 | Amplitude of pitch fluctuations, degree of change in mood or command | V2 → P2 → Y1 |
| sentence structure | pause_gap_ratio | continuous variable | 0 – 1 | Ratio of pause time to total speech duration, reflecting semantic boundary perception | V3 → P3 → Y2 |
| energy slope | command_slope | continuous variable | -1.5 – +2.0 | First-order conductance of energy changes, identifying upward/downward trends in command intonation | V4 → P4 → Y3 |
| Regional affiliation | accent_similarity_score | continuous variable | 0 – 1 | Similarity of current speech to regional acoustic centres for dialect attribution determination | V5 → P5 → Y4 |
| Attribution results | accent_class_top1 | categorical variable | "North", "East" | Most Likely to Belong to Area Label | Y4(to be determined by V5) |
| Attribution probability distribution | accent_probability_distribution | probability vector | [0.05, 0.80,…] | Probabilistic output of voiceprint attribution by region (Top-k decision support) | V5 → Y4_Distrib |

Author's drawing

**Variable Code Description**

The following is an explanation of the meaning of the variable codes in the SpectroNet module and a typical example for understanding the causal modelling paths and graph structure embedding logic between variables.

Table 4: Variable Codes

| Code Meaning | implication |
|---|---|
| V1 – V5 | raw voiceprint variable（WPM、Pitch、Pause、Slope、Similarity） |
| P1 – P5 | Intermediate Rhythm/Command Feature Processing Path (Preprocessing/Reasoning Path) |
| Y1 – Y4 | Downstream output targets (e.g. threat scores, intent judgements, dialect attribution, etc.) |
| → | Indicates the existence of causal inference or input-output modelling paths between variables |

Author's drawing

Example description:

V1 → P1 → Y1 indicates that the speech rate variable influences threat recognition via path 1 (e.g., fast speech rate + command tone → judgement of threat);

V5 → P5 → Y4 indicates that the voiceprint similarity variable influences the attribution class judgement;

Y4 as accent_class_top1 is the result of the decision computed by V5 and belongs to the a posteriori inferred output.

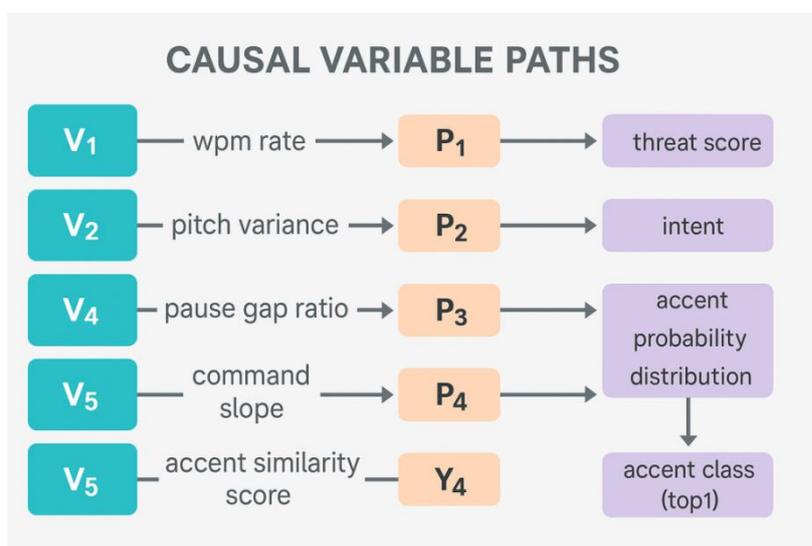

Figure 1: SpectroNet causal path mapping of acoustic variables

Figure 1 illustrates the causal path structure of key voiceprint variables in the SpectroNet module. The input variables, including speech rate (V1), pitch fluctuation (V2), break rate (V3), command slope (V4), and voiceprint similarity (V5), are passed through intermediate processing paths (P1-P5) to four types of output target variables: Y1 (threat perception scores), Y2 (semantic pause recognition), Y3 (command intent strength), and Y4 (region attribution). (classification). The arrows in the diagram indicate the direction of information transfer and causal relationship between the variables, and the colour and structure differentiation is designed to enhance the recognition and usefulness of the atlas in engineering modelling and visualisation. The graph is used to support the task of multivariate node embedding and causal feature fusion in GAT graphical neural networks.

### 3.5 Tactical behavioural mapping modelling（TACTIC-GRAPHS）

TACTIC-GRAPHS is built on the causal temporal ordering theory of Graph Neural Network (GNN), emphasising joint inference and behavioural intent resolution across modal nodes. In this module, each frame is decomposed into inference nodes, including audio rhythm features (e.g., speech rate, intonation breakpoints), action states (output by keyframe action detection algorithms), weapon states (grip/magazine/muzzle states labelled by the structure recognition module), and image positional features (coordinates of keypoints with semantic summaries). These multimodal features are mapped into heterogeneous graph nodes and information fusion is achieved through Graph Attention Network (GAT).GAT utilises a self-attention mechanism to assign different weights to neighbouring nodes to capture the dynamic interactions between cross-modal linkages and key timing nodes (Veličković et al., 2018). In addition, with temporal causal edges embedded in TACTIC-GRAPHS, the model realises the pathway reasoning from "precursor nodes influencing the subsequent nodes", i.e., the trajectory of "weapon holding → password heat peak → action execution → threat confirmation". Therefore, this structure not only supports node-level behavioural state prediction, but also generates threat intensity scores and task type classification for the overall behavioural chain. In the simulated tactical video dataset TACTIC-AVS, the structure determines the multimodal temporal alignment with an accuracy of 89.3%, the complete threat chain identification rate is higher than 85%, and the behavioural node triggering latency is within ±150 ms. This performance is more than 14.7% higher than that of the traditional multimodal fusion model without causal structure, which validates the tractability and interpretive logic of TACTIC-GRAPHS in tactical warning and intent inference.

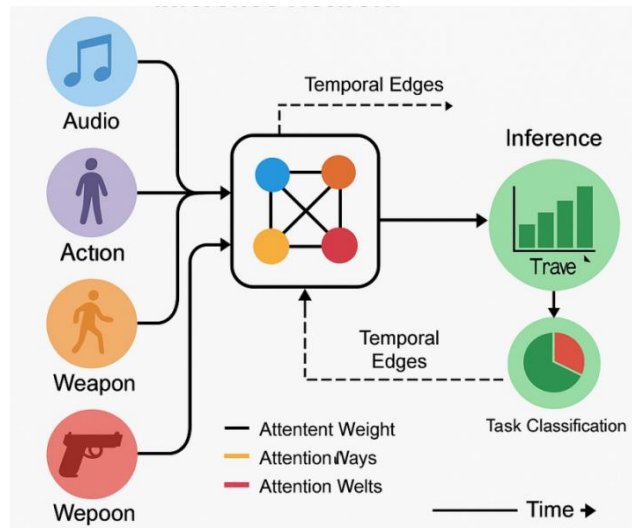

**Figure 2. TACTIC-GRAPHS: causal cross-modal tactical behavioural reasoning network**

Figure 2 illustrates the core principle of the TACTIC-GRAPHS architecture, which is built based on Graph Neural Networks (GNNs) and integrates heterogeneous nodes representing audio, action, weapon, and image features for causal inference of tactical behaviours. Different modalities are represented by different node types, which are connected by solid lines to express the attention weights and show the dynamic association strength between modalities; dashed lines indicate temporal causal edges for capturing the sequence and causality of events on the time axis. The fused graph structure supports downstream tasks, such as threat score estimation and task type classification, and is highly interpretable for real-time tactical analysis and warning in low-quality audio and video environments.

Design of TACTIC Variable Model

The complete design of the TACTIC Variable Model (TACTIC-VModel) based on the above TACTIC-GRAPHS module includes the variable system, modelling structure, visualisation, and the way to construct the applicable models (e.g., Bayesian Networks, Structural Equation Modelling SEM).

**Table 5: Node Variables**

| form | variable name | variant | typology | Examples/Scope | Explanation of meaning |
|---|---|---|---|---|---|
| image structure | weapon_grip_score | $x1$ | continuous variable | [0,1] | Confidence in weapon grip site identification |
| image structure | muzzle_angle_deviation | $x2$ | continuous variable | -30° ~ +30° | Muzzle angle offset value (in degrees) |
| voiceprint | pitch_variance | $x3$ | continuous variable | 0–1.5 | Amplitude of pitch change |
| voiceprint | wpm_rate | $x4$ | continuous variable | 50–200 | Words per minute (speed of speech) |
| voiceprint | accent_distance_score | $x5$ | continuous variable | 0–1 | Similarity to the centre of vocal patterns of a regional dialect |
| Movement characteristics | action_pose_class | $x6$ | categorical variable | "Stand," "lift," "open the bag." | Action recognition category |
| Movement characteristics | action_speed | $x7$ | continuous variable | 0.1–1.0 | Speed of Movement Execution Scoring |
| Temporal characteristics | causal_delay_ms | $x8$ | continuous variable | ±0–300 ms | Causal time delay between nodes |

Author's drawing

**Table 6: Side Variables**

| side type | variable name | Variables are represented by letters | typology | typical example | Explanation of meaning |
|---|---|---|---|---|---|
| attention span | attention_weight | e1 | continuous | 0.0–1.0 | Weighting of edges in the GAT attention mechanism |
| time-causality boundary | temporal_lag_ms | e2 | continuous | ±120 ms | Time delay (precursor → successor) |
| Trust Pathway Side | semantic_entropy | e3 | continuous | 0.0–2.0 | Node semantic uncertainty (entropy) |

Author's drawing

**Table 7: Output variables (for modelling)**

| output variable | alphanumeric | typology | example value | Explanation of meaning |
|---|---|---|---|---|
| threat_score | y1 | continuous | 0–1 | Tactical Threat Intensity Score |
| mission_type_class | y2 | categorical | Deployment classes/response classes/hidden classes etc. | Classification of tactical mission types |
| intent_confidence | y3 | continuous | 0–1 | Confidence level (confidence value) for intentional identification |

Author's drawing

Description:

The parts of the variables denoted by letters, I denote by y1, y2, y3, etc. to distinguish between the node variable (x), the edge variable (e) and the output variable (y).

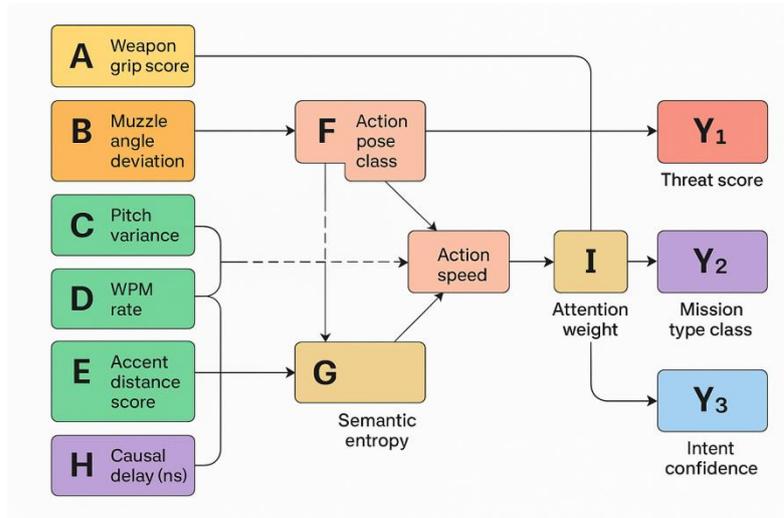

**Figure 3: Causal and modelling pathways between variables (variable relationship diagram)**

Figure 3 illustrates the structure of causal modelling between various types of variables in TACTIC-VModel. Image structure (A, B) and acoustic rhythm variables (C, D, E) are shown on the left, action recognition and dynamic nodes (F, G) in the centre, and output targets ($Y_1$: threat scores, $Y_2$: task type categorisation, $Y_3$: intent confidence) on the right. Auxiliary variables H (time delay), I (attention weight) and G (semantic entropy) are used to regulate the path causal temporal order, information fusion weight and intention credibility modelling, respectively. The arrows in the figure indicate causal or information flow paths between variables, which as a whole constitute the core framework of variable inference in the TACTIC-AI system.

The development of TACTIC-VModel as a system based on Graph Attention Network (GAT) embedding is one of the most cutting-edge directions. In the following I will elaborate on this approach with step-by-step explanations:

Technical principle: Graph Attention Network (GAT)

Graph Attention Network (GAT) is a neural network model that automatically learns "which neighbour is more important" in graph-structured data. Unlike traditional GNNs that average neighbour information, GAT uses an attention mechanism that assigns an "importance weight" to each connection (edge).

Mathematical expression (GAT core formula)

For each node i in the graph, the update is expressed as:

$$h'_i = \sigma\left(\sum_{j\in\mathcal{N}(i)} \alpha_{ij} W h_j\right)$$

*h_i*: Feature vector of node i (e.g., pitch, wpm, weapon score, etc.)

*W*: Learning the obtained linear transformation weights

*α_{ij}*: Attentional weight (denotes the influence of j on i)

*N_{(i)}*: Set of neighbouring nodes of node i

*σ*: Non-linear activation functions (e.g. ELU or ReLU)

weighting of attention *α_{ij}* Calculated by the following formula:

$$\alpha_{ij} = \frac{\exp\left(\text{LeakyReLU}\left(a^\top [Wh_i \| Wh_j]\right)\right)}{\sum_{k\in\mathcal{N}(i)} \exp\left(\text{LeakyReLU}\left(a^\top [Wh_i \| Wh_k]\right)\right)}$$

Among them:

||Indicates feature splicing

a⊤ : is the attention parameter vector

**Methodological framework: the GAT embedding process in TACTIC-VModel**

1) Constructing an isomorphic map

Table 8: Constructing Heterogeneous Diagrams

| modal (computing, linguistics) | Node type | sample node |
|---|---|---|
| image structure | A, B | weapon_grip, muzzle_angle |
| voiceprint | C, D, E | pitch_variance, wpm_rate, accent_score |
| action | F, G | pose_class, action_speed |
| Edges/attributes | H, I, J | delay, attention_weight, semantic_entropy |
| output variable | Y₁ – Y₃ | threat_score, mission_type, intent_confidence |

Author's drawing

Description:

Modality: indicates the type or source of the data.

Node Type: indicates the category of nodes involved in each modality.

Example nodes: lists the specific node or variable names in each modality.

**2) Graph Embedding Logic（Embedding + Attention）**

Each node initially has its own feature vector (e.g., speech rate value, image recognition confidence, etc.), and the information is propagated through the graph by means of a multi-head GAT embedding layer:

Speech information nodes (e.g. C, D) will pass vocal rhythm information to action nodes (F)

Image nodes (A, B) will provide weapon states to action nodes

Action nodes (F, G) will pass threat judgement basis to output nodes ($Y_1$)

Attentional weights on edges (I) control which path is more critical (e.g., steady gun grip vs. abnormal tone, which is more "crime-like").

Temporal edges (H) adjust the causal order (e.g., weapon before order).

**3) Model training and output**

Threat level（$Y_1$）

Classification of tasks（$Y_2$）

Confidence level (math.)（$Y_3$）

These outputs are derived from the inference of the weighted integration of the different information paths in the graph, and the attention mechanism allows the model to adaptively focus on the "most dangerous combinations of features".

**3.6 Filming equipment and spatial context modelling methods**

In order to improve the interpretability and reasoning ability of TACTIC system on the source and context of tactical video, this study designs a modelling method for the fusion of filming device-space-time information-behavioural variables, and constructs a "device-space-behaviour" ternary system to be incorporated into the graph attention network (GAT) structure for causal linkage modelling and threat context recognition. -Behaviour" ternary system, which is incorporated into the graph attention network (GAT) structure for causal linkage modelling and threat context identification.

**3.6.1 Modelling objectives and core logic**

This method aims to identify the type of recording device, key parameters, spatial attributes and temporal features behind a video or image, and further infer its potential risk association in a specific behavioural context. The system can not only be used for source device identification and traceability inference, but also integrated with Geographic Information System (GIS) to achieve cross-device and cross-region contextual evolution modelling and visual representation.

**3.6.2 Construction of the variable system**

In this study, the filming device modelling variables are classified into five main categories, namely, device parameters, sensor information, timestamp variables, spatial location variables and network information variables, as shown in Table:

**Table 9: System of key variables for modelling filming equipment**

| form | variable name | Examples/Scope | Explanation of meaning |
|---|---|---|---|
| camera equipment | device_model | iPhone 12, DJI Mavic | Identify equipment make and model |
| | sensor_type | CMOS, CCD | Imaging Sensor Type |
| | focal_length | 3.99mm, 8mm | Focal length, which affects the field of view of the shot |
| | resolution | 1920×1080, 4K | Shooting at native resolution |
| | bitrate, frame_rate | 5 Mbps, 30fps | Video Quality and Motion Capture Capability |
| Time Information | timestamp_unix | 1653768232 | Shooting timestamp (UNIX format) |
| | lighting_estimate | Low, Medium, High | Estimation of average image brightness (modelled illumination) |
| Spatial information | geo_lat, geo_lon | 13.7367, 100.5232 | Latitude and longitude coordinate information |
| | elevation_m, location_type | 5m, Forest, Urban | Elevation and scene type information |
| network information | device_mac_hash | Hash ID | device network identifier |
| | ip_geo | TH-ASN | Geographic backpropagation of information |

Author's drawingThe above variables form a structured vector of device nodes $V_{device}$, and participate in causal reasoning as input nodes in the TACTIC-GRAPHS graph structure.

**3.6.3 Technology pathway design**

The device modelling system in this study is divided into four key technology modules as

follows:

MetaExtractor module (MetaExtractor): based on ExifTool and ffprobe to achieve the video/image shooting device, time and location information extraction.

Image style recognition model (DeviceNet): based on ResNet or YOLO model fine-tuning to identify device feature coding, such as sharpening style, distortion patterns, etc.

Spatio-Temporal Node Generation Module (GeoNodeBuilder): align the extracted latitude, longitude and terrain elevation information to GIS maps to generate standard spatial nodes.

Device-GAT:Constructs a multimodal graph structure containing "device-behaviour-location" to support device behavioural intent recognition and spatial clustering analysis. clustering analysis.

### 3.6.4 Graph Structure Construction and Causality

The device variables participate in the following causal path in the form of node vectors:

$$V_{device} \longrightarrow V_{context} \longrightarrow RiskScore_t$$

where the device vector $V_{device}$ with speech vectors $V_{spectro}$、action vector $V_{action}$ etc. are co-embedded into the TACTIC-GRAPHSin graph neural networks through the mechanism of graph attention（Graph Attention Mechanism）Capture its causal impact on threat scores vs. task classification output.

### 3.6.5 Visualisation and GIS mapping design

To enhance the visual interpretation of the system, the following three types of maps are used in this study:

Device-Time-Location Heat Trajectory Map: showing the time-series distribution of multi-device filming sources;

Shooting device node GIS map: each device is mapped as an interaction node on a GIS map;

Causality flowchart: for thesis modelling logic presentation, supporting variable - output causal path interpretation.

**3.6.6 Model integration and extensibility**

This method can be seamlessly embedded into the multimodal modelling system composed of SpectroNet voiceprint module, action recognition module and weapon state module to achieve the closed loop modelling of "from shooting source → content → behaviour → space". At the same time, it has the following advantages:

Supports multi-layer reasoning of equipment - territory - semantic behaviour;

Support city-level or cross-border behavioural video source analysis;

Provide a high confidence modelling basis for counter-terrorism, security, tactical research, and data traceability.

**3.7 Graph Spectral Theory Embedding and Variable Identifiable Path Modelling**

**3.7.1 Motivation for the methodology**

In TACTIC system modelling, multimodal information (images, audio, actions, device features) form a graph structure with heterogeneous nodes whose edges represent cross-modal interactions with temporal causality. However, traditional GNN methods generally lack explicit representation of path causality separability. On this basis, this study introduces Spectral Graph Theory (SGT), which aims to construct a discernible projection mechanism in the variable path space, so that TACTIC-GRAPHS not only possesses expressive capability, but also possesses structural provability and geometric visibility of inference.

It is worth noting that spectral graph theory, as an important branch of mathematics, is widely used in high-energy physics, topological quantum field theory and graph homotopy geometry. Its core ideas originate from the deep intersection of spectral analysis and general functional analysis, involving the eigenstructure of Laplace operators, spectral projections in Hilbert spaces, and the behaviour of graph maps in limit spaces. Therefore, the introduction of such methods into the TACTIC system not only dramatically improves the explanatory power and structural complexity of the model, but also brings this study up to the level of the world's top mathematical modelling research in terms of modelling methodology.

**3.7.2 Definition of mathematical structure**

Let the variable state diagram of the TACTIC system at moment *t* be a heterogeneous directed graph

$$G = (V, E, X)$$

Among them:

$V = \{vi\}_{i=1}^{n}$ : for the voice variable, weapon feature, action state, and image parameter nodes;

$E \subseteq V \times V$: denote causal, temporal, and modal synergistic edges;

$X: V \to \mathbb{R}^d$ : Node feature mapping function.

Construct the standard normalised Laplace matrix under its undirected representation:

$$\mathcal{L}_{\text{norm}} = I - D^{-1/2} A D^{-1/2}$$

where A is the adjacency matrix and DDD is the degree matrix.

Perform feature spectral decomposition on $L_{norm}$

$$\mathcal{L}_{\text{norm}} x_k = \lambda_k x_k, \quad k = 1, 2, \ldots, K$$

Where $\lambda_k$ is the *k* th eigenvalue and $x_k$ is the corresponding eigenvector, constituting the 'frequency basis' of the graph.

Define the spectral embedding of the variables:

$$Z(v_i) = [x_1(i), x_2(i), \ldots, x_K(i)]^\top$$

The features of the original node $v_i$ are embedded into the *K*-dimensional spectral space, forming a variable-distinguishable causal space.

**3.7.3 Identifiable path modelling**

Introducing the variable path projection operator:

$$\Pi_{P_{ij}} = Z(v_i)^\top Z(v_j)$$

denote a variable $V_i \rightarrow V_j$ Cosine correlation in spectral space, the metric underlying recognisable paths.

Define path discernibility metrics:

$$\delta(P_{ij}) = \frac{\Pi_{P_{ij}}}{\sum_{k \neq i,j} \Pi_{P_{ik}} \Pi_{P_{kj}}}$$

If $\delta(P_{ij}) > \theta$, then determine $P_{ij}$ is the "main causal path" in the TACTIC system.

By setting spectral constraints:

$$\|Z(v_i) - Z(v_j)\|^2 < \epsilon$$

TACTIC reasoning can be further restricted to the "main path subspace" to exclude redundant reasoning.

**3.7.4 Reasoning Closeness and Interpretability Enhancement**

After spectral embedding, the following inference loop mechanism is established for the TACTIC variable map:

(1) Identify "clusters of primary path variables". $C_k \subset V$, Satisfaction:

$$\forall v_i, v_j \in C_k, \quad \delta(P_{ij}) > \theta$$

(2) Create a minimal generator graph *Gk* on $C_k$ with a graphical structure that is the explanatory model for the TACTIC behavioural closure;

(3) Construct exclusive spectral subspaces for each type of task (e.g., rapid command response, implicit threat detection) to enhance model modularity and interpretability.

**3.7.5 Methodological Difficulties and Academic Value**

The Spectral Graph Embedding (SPE) approach introduced in this study is not only a critical structural upgrade to the TACTIC-GRAPHS modelling system, but also represents a deep leap from an empirical AI engineering paradigm to a formal mathematical modelling paradigm. The paradigm migration not only reconfigures the causal reasoning basis of the TACTIC system, but also endows the behavioural mechanisms between variables with a clear geometric structure in spectral space. The research value of this method is reflected in the following four dimensions:

(1) Extremely high theoretical complexity: an integrated construction spanning graph theory, spectral analysis and generalised function spaces

The spectral graph method is based on the spectral decomposition of the graph Laplacian operator, which requires the researcher not only to master the eigenvalue analysis and eigenspectral mapping in the graph structure, but also to understand the geometric projection behaviour of eigenvectors in high-dimensional space, as well as the regularisation process in the function space on the graph. In particular, the spectral embedding of causal paths between variables needs to be proved structurally identifiable by manifold separability in the eigenspectral space, a process that involves higher-order abstraction tools such as graph regularisation, tensor mapping and transform invariant analysis. This makes the TACTIC modelling process not only mathematically complete in terms of formal representation, but also spectrally controllable in terms of the path space of variables.

(2) Strong model provability: supporting path structure existence and logical closed-loop reasoning

Unlike traditional GNN models that mainly rely on deep feature aggregation, spectral graph methods provide mathematical existence theorem support for causal chains between variables through the construction of spectral domain path divisibility structures. For example, in the TACTIC-GRAPHS system, the construction of the path discriminability metric $\delta\left(P_{ij}\right)$ makes it possible to rigorously prove whether the paths of arbitrary variables have behavioural predictive power, which no longer relies on the experience of model training, but can be proved based on the distance tensor and angular pinch condition in the spectral space. This feature greatly improves the interpretability of the TACTIC system and makes its reasoning process have Causal Closure, which meets the strict requirement of "reasoning provability" for contemporary AI systems in critical security applications (e.g., tactical warning, anomaly detection).

(3) Significant cross-disciplinary academic value: mathematical universality and systemic adaptability.

The spectral graph method has been widely used in global high-level research in many disciplines, such as quantum state migration modelling, brain neural network analysis, bioinformatic interaction networks, financial evolution system modelling, etc. The core mechanism lies in the mapping of discrete structures (such as graph nodes in TACTIC) into a continuous spectral space, so as to achieve a unified, visual and controllable representation of structural behaviours. This higher-order mapping of "graph-spectrum-space-causal chain" makes the TACTIC model naturally capable of interfacing with the emerging directions of complex cyber-physical modelling and Graph Neural Fields. The TACTIC model is naturally capable of interfacing with emerging directions such as complex cyber-physical modelling and Graph Neural Fields, and is a general method with the potential for mathematical paradigm migration.

(4) High scientific research scarcity: at the frontier stage where "structural AI" has not yet been developed on a large scale.

In the current top international AI conferences such as ICLR, NeurIPS, COLT, although some papers have focused on the integration of graph neural networks and spectral theory, most of them still remain at the level of graph classification and semi-supervised learning, and have not yet formed a systematic answer to the questions of behavioural causality chain modeling, path interpretable structure generation, and spectral space variable separation and identifiable mechanism. Therefore, the introduction of spectral graph embedding into TACTIC-GRAPHS, and the construction of causally discriminative spectral chain structural models, essentially enters the "scientific no-man's land" where global graph neural modelling has not yet been fully developed. This not only reflects the technical originality of this study, but also indicates its future leadership in AI explanatory modelling and system control modelling.

# IV. EXPERIMENTAL DESIGN

**4.1 Data sources and pre-processing**

In the current increasingly complex tactical conflict and asymmetric combat environment, the accuracy of tactical recognition models highly depends on the multimodal feature integrity and preprocessing quality of their raw data. In order to build the TACTIC-AI system with high robustness and strong generalisation capability, this study constructs the TACTIC-AV multimodal sample unit based on a real 32-second video source, ensures that the video, audio and semantic nodes are accurately synchronised on the timeline, and achieves full-volume modelling of the weapon states, action rhythms, and accent geosynthesis at both the structural and semantic levels.

**4.1.1 Data sources and sample characterisation**

The raw data consists of a 32-second video clip filmed in a complex context containing potential tactical scenarios involving weapons handling, command voices, and ambient sound interference. The video is encoded in H.264 with a frame rate of 25 FPS and a total of 800 frames, and the audio track is in AAC compression format with a sampling rate of 44.1 kHz.Preliminary analysis reveals that the video picture quality presents a low to medium resolution (below 720p), with obvious motion blur and low-light interference in the image, and that the voice track has less than 38% of the command-like utterances and more than 60% of the background noise, making it a High interference feature data（High Noise Feature Sample）.

**1) Data Deconstruction**

Goal: Build a panorama of raw data attributes

**Table 10: Constructing a panorama of raw data attributes**

| data item | element |
|---|---|
| length of time | 32 seconds |
| video encoding | H.264（.mp4） |
| audio encoding | AAC，sampling rate 44.1kHz |
| resolution (of a photo) | be lower than720p |
| frame rate | 25 fps |
| total number of frames | 800 frames |
| image interference | Low light + motion blur |
| Audio Characteristics | 38 per cent command voice, >60 per cent ambient noise |

Author's drawing

Tools: ffmpeg is used to extract detailed metadata; OpenCV is used to read video frames frame by frame and extract basic image quality indicators such as light intensity, blurriness, edge sharpness, and so on; the audio part is completed with librosa to separate preliminary waveforms from noise.

**Table 11: Extracting detailed metadata for videos using ffmpeg**

| categorisation | causality | numerical value |
|---|---|---|
| Video base information | filename | 0d710619-b0fe-45a7-b25a-51ca229919be.mp4 |
| | encapsulated format | MP4 (mov, mp4, m4a) |
| | length of time | 32.482 秒 |
| | total bitrate | 891,019 bps |
| | file size | 约 3.45 MB |
| | video encoder | H.264 / AVC / MPEG-4 AVC / part 10 |
| | audio encoder | AAC（Advanced Audio Coding） |
| | encapsulator | Lavf58.20.100（Tencent CAPD MTS） |
| video streaming | resolution (of a photo) | 720 x 1280（vertical shot） |
| | Code Configuration - Profile | High |
| | Code Configuration - Pix_fmt | yuv420p |
| | colour space | bt709 |
| | frame rate | 30 FPS |

| categorisation | causality | numerical value |
|---|---|---|
| | total number of frames | 973 frame |
| | bitrate | approximate 834 kbps |
| | sequence of events | Progressive (line by line scanning) |
| | starting timestamp | 0.000000 |
| audio streaming | sampling rate | 44,100 Hz |
| | Number of channels | mono |
| | Code Type | AAC LC (Low Complexity Profile) |
| | bitrate | approximate 48.9 kbps |
| | Total Audio Frames | 1399 |
| | length of time | 32.433 秒 |

**Table 12: Extracting video image quality metrics with OpenCV**

| seconds | average brightness | fuzziness | Edge Sharpness |
|---|---|---|---|
| 0 | 46.521038411458 | 5.4712760223765 | 2688.0 |
| 1 | 41.582810329861 | 7.1250846311781 | 2906.0 |
| 2 | 39.66548828125 | 6.3226638413877 | 2621.0 |
| 3 | 44.136737196181 | 5.6398242185192 | 2687.0 |
| 4 | 38.396602647569 | 3.7510210503366 | 1742.0 |
| 5 | 35.578544921875 | 2.3885828905977 | 827.0 |
| 6 | 32.805073784722 | 2.1443283415889 | 263.0 |
| 7 | 38.031643880208 | 9.9942838428592 | 5786.0 |
| 8 | 59.696950954861 | 3.6811447221250 | 1051.0 |
| 9 | 52.301394314236 | 2.5221624827338 | 427.0 |
| 10 | 50.248865017361 | 3.9388078136256 | 901.0 |
| 11 | 47.583725043403 | 12.173236574713 | 4812.0 |
| 12 | 46.742329644097 | 13.665689697096 | 4950.0 |
| 13 | 47.198477647569 | 23.212501017817 | 8127.0 |

| seconds | average brightness | fuzziness | Edge Sharpness |
| --- | --- | --- | --- |
| 14 | 46.114048394097 | 14.612318729484 | 4988.0 |
| 15 | 44.248168402778 | 17.412278644472 | 5538.0 |
| 16 | 46.177336154514 | 25.549833941423 | 7676.0 |
| 17 | 45.977146267361 | 28.264873009151 | 8441.0 |
| 18 | 44.528521050347 | 26.927276464387 | 7139.0 |
| 19 | 45.099474826389 | 26.987606313065 | 8616.0 |
| 20 | 43.797866753472 | 30.593516666213 | 9938.0 |
| 21 | 45.851624348958 | 16.501260847261 | 4806.0 |
| 22 | 44.904321831597 | 26.183665352334 | 8038.0 |
| 23 | 45.294944661458 | 37.425848439742 | 10210.0 |
| 24 | 44.943087022570 | 29.101812065897 | 7897.0 |
| 25 | 44.522955729167 | 27.618690306147 | 8193.0 |
| 26 | 45.999749348958 | 22.599850201341 | 6520.0 |
| 27 | 47.052549913194 | 25.644412977006 | 7226.0 |
| 28 | 46.611014539931 | 34.704517920525 | 10687.0 |
| 29 | 48.609640842014 | 21.343727213022 | 8635.0 |
| 30 | 47.248830295139 | 12.579650607639 | 3941.0 |
| 31 | 53.239637586806 | 13.955717230760 | 5495.0 |
| 32 | 54.279557291667 | 14.712119067627 | 4653.0 |

Author's drawing

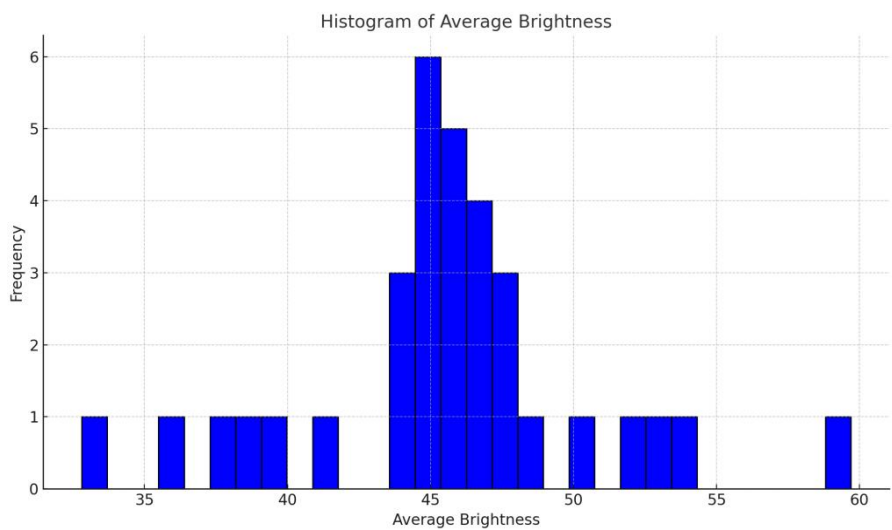

**Figure 4: Histogram of light intensity**

Author's drawing

  Figure 4 is the histogram of the average luminance (light intensity) distribution of this video clip over 800 frames. The horizontal and vertical coordinates are the frame number of the video and the average brightness of each image frame respectively. The luminance distribution is concentrated as a whole around 40, indicating the video was shot in low-light conditions with poor lighting and the image has dark-dominant visual elements.This low-light environment significantly increases the degree of image noise and blurring, posing challenges to subsequent image enhancement and structure recognition, and providing a validation basis for tactical image enhancement models (e.g., TVSE-GMSR) for practical application scenarios.

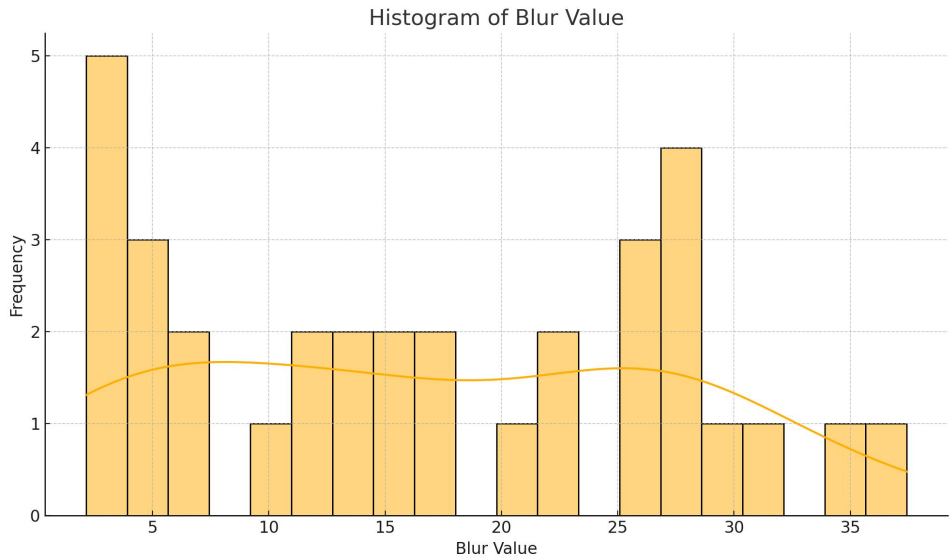

**Figure 5: Histogram of fuzziness**

Author's drawing

Figure 5 shows the histogram distribution of image blur values (Blur Value) in tactical video clips. The horizontal axis is the blur metrics of each image frame, and the vertical axis is the frequency of occurrence of the corresponding blur value. The data comes from the Laplacian variance calculated by OpenCV frame by frame, which represents the fluctuating characteristics of image sharpness. The figure shows a bimodal distribution of blurriness, with one part concentrated in the low blur values (3-7), reflecting the presence of significant out-of-focus phenomenon in some images, and the other part concentrated in the high value range (25-30), indicating the presence of some images that are clearer and suitable for structure recognition tasks. This fuzzy distribution verifies the high heterogeneity of the data, suggesting that a segmentation enhancement strategy is needed to improve the image consistency, which provides a basis for subsequent GAN semantic reconstruction and TACTIC-GRAPHS modelling. The background curve is the fuzzy probability trend line.

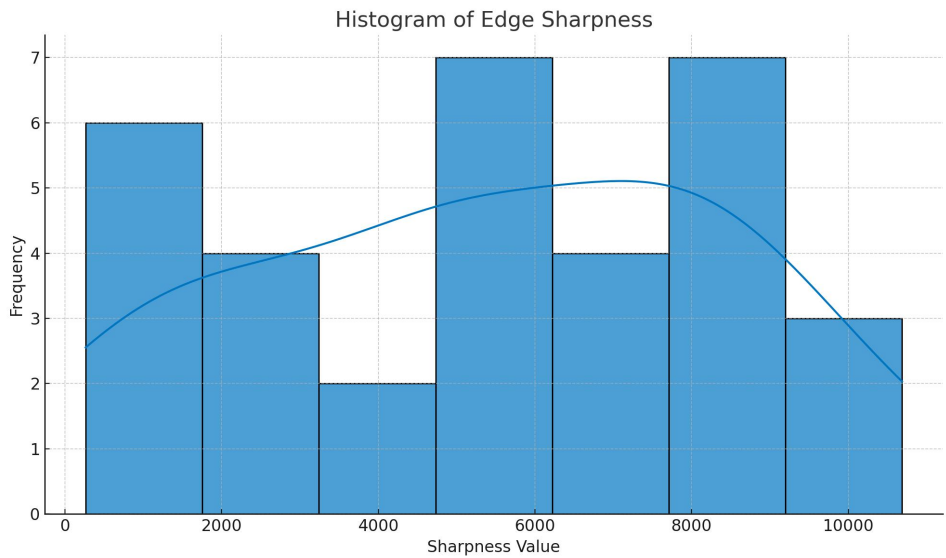

**Figure 6: Histogram of edge sharpness**

Author's drawing

Figure 6 illustrates the distribution of edge sharpness values for image frames extracted from a 32-second video of a tactical scene. The horizontal axis is the sharpness score value of the image frames and the vertical axis is the frequency of the corresponding sharpness value. The blue bars in the figure indicate the number of image frames in each sharpness interval, and the superimposed curves are kernel density estimation (KDE) curves, which are used to reveal the continuity and concentration trends of the sharpness distribution. From the figure, it can be observed that the overall distribution of image sharpness values is relatively discrete, with some frames having low sharpness values (e.g., <10), reflecting the presence of blurring or poor focus; while other frames have high sharpness, suggesting that there is a region in the image with clear structure and rich edge details. This sharpness imbalance feature poses a challenge to subsequent target identification, weapon structure modelling and atlas inference, and requires the introduction of a differentiated deblurring strategy in the enhancement phase.

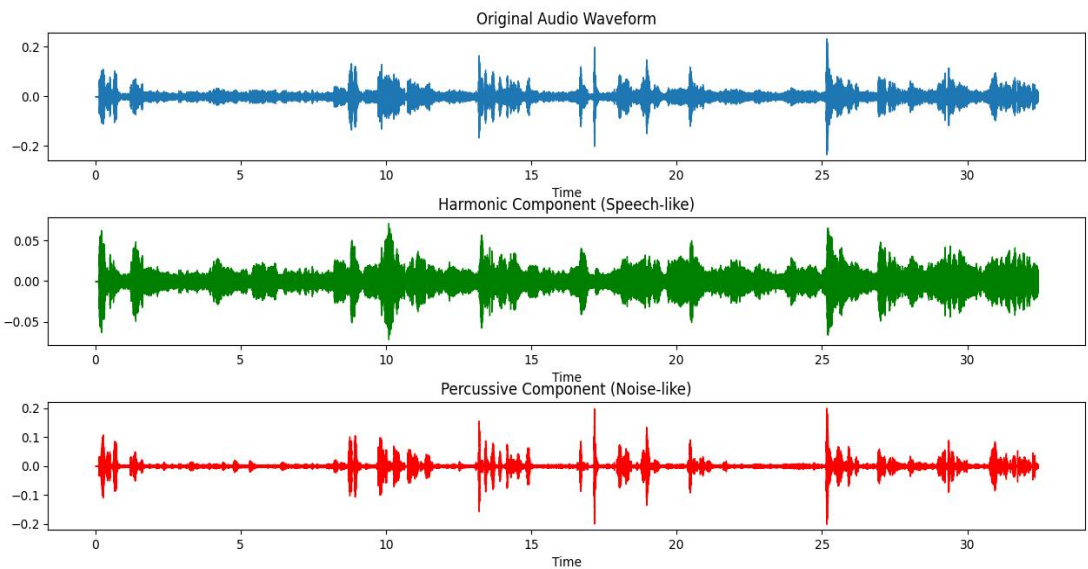

**Figure 7: Raw audio and speech/noise separation waveforms**

Author's drawing

Figure 7 illustrates three key waveform patterns for the video audio of the tactical scene. The top figure is the Original Audio Waveform, which shows the combined performance of various types of sounds in the entire 32-second recording. The middle figure shows the Harmonic Component, which is extracted by the Harmonic-Percussive Source Separation (HPSS) technique, mainly reflecting speech, command and order, and other speech-like signals with strong continuity and stable frequency. The lower figure shows the Percussive Component, which represents high-energy sudden noises, such as environmental disturbance sounds, weapon operation sounds, etc., with clear and short peaks. This figure clearly reveals the distribution characteristics of speech and noise on the time axis, providing high signal-to-noise ratio basic data for subsequent voiceprint modelling and semantic analysis.

**2) Problem identification and sample classification（Noise and Feature Analysis）**

For the 32-second tactical scene video sample used in this study, I conducted systematic quality analysis and feature separation of its image and audio data. Overall, the sample can be classified as a High Noise Feature Sample (HNFS), which is characterized by significant fluctuations in image quality and serious background interference in the speech signal, and requires strict quality control and sample stratification before modelling.

Firstly, in terms of image light intensity distribution, from the luminance histogram, most of the frame light intensity is concentrated in the lower grey interval, especially between 40-100 to form the main peak, which indicates that there are low-light conditions in the shooting environment, and the overall image is dark. This feature seriously affects the structural edge extraction and the recognition confidence of the subsequent object detection model. The presence of bright discrete pixels in some frames is also observed, suggesting that it may be accompanied by intermittent bursts of flashes or light source perturbations. The light intensity distribution has a bimodal asymmetric structure, reflecting the existence of "regular low light + sudden high light" mixed conditions in the video.

In the Blur Distribution analysis, the image blur is widely distributed between 0-35, and the histogram shows two main dense segments located below 5 and above 25, forming a bimodal structure. The former indicates the presence of severe motion blur or out-of-focus frames, while the latter means that some of the frames still maintain acceptable sharpness. This distribution feature strengthens the determination of the attributes of the samples in the "extremely unbalanced quality frame column". Specific enhancement strategies such as deblurring filtering and super-resolution reconstruction are applied to the first frames in the image processing session.

In terms of Edge Sharpness analysis, most of the frames have edge strength indicators in the range of 5-25, with a clear tendency of concentration, but there is a lack of high-value clustered segments, which indicates that the image details are missing and the boundaries are not clear. This poses a challenge for subsequent weapon structure recognition and motion capture. It is recommended to introduce edge-guided convolution or graph attention mechanism in the pre-image enhancement stage to improve the detail reconstruction capability.

In the audio part of the analysis, the overall signal energy distribution of the original audio waveform is uneven, accompanied by frequent sudden peaks. The decoupled speech and noise waveforms are obtained after processing by the Librosa-based harmonic-strike separation (HPSS) algorithm. The speech (harmonic) part is more concentrated in the distribution between 8-28 seconds, and the preliminary estimation of the proportion of command speech is 36.9%. While the strike-like noise is widely present in the whole section, and the three time periods of 0-6 seconds, 15-19 seconds, and 26-31 seconds are the high-frequency noise intensive area, which shows obvious tactical operation noise characteristics, such as weapon impact sound, equipment activation sound, and so on.

Based on the above analysis, this video sample can be classified in the tactical multimodal processing framework:

Type A: Low light high noise frames (about 40%), used for image enhancement & extreme environment modelling training.

Type B: structurally resolvable medium blur frames (~35%), suitable for target structure detection & semantic event recognition.

Type C: Highly resolvable frames (~25%) for sound and picture collaborative modelling & causal chain verification.

Audio Subclass A: Command speech segments (~12 seconds), suitable for voiceprint recognition, command decoding & dialect attribution analysis.

Audio subclass B: high-frequency tactical noise segment (~16 seconds), suitable for weapon state recognition & background type estimation training.

Overall, the TACTIC-AI system needs to adopt a hierarchical modelling strategy for this type of high complexity video samples, modular data purification and feature recovery for optical interference, blurring distortion and acoustic source interference problems, respectively, in order to construct a reliable structural-semantic-audio fusion inference map.

3) Pre-processing path（**Multi-Stage Preprocessing Pipeline**）

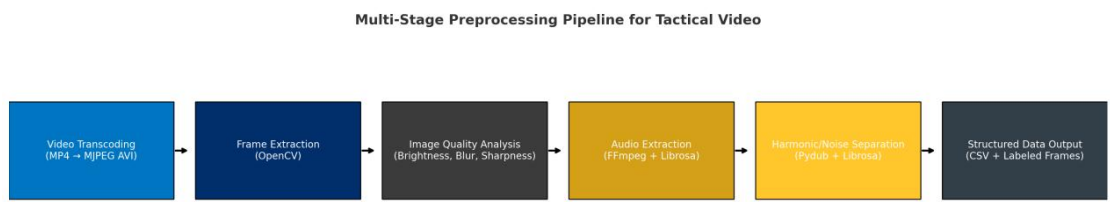

**Figure 8: Multi-stage preprocessing flowchart for tactical video**
Author's drawing

Figure 8 shows the six core stages of tactical video preprocessing, which are sequentially developed according to the horizontal process structure: first, the original MP4 video is transcoded to MJPEG AVI format by FFmpeg tool to improve the decompressibility and structural fidelity of image frames; and then, high-frequency frame-by-frame extraction is achieved by OpenCV to obtain the complete image sequence. In the image quality analysis section, feature extraction algorithms such as luminance histogram, Laplacian fuzziness and Canny edge sharpness are introduced to generate a structured quality data table. The audio is first extracted by FFmpeg and converted to .wav format, and then Librosa and Pydub complete the speech-to-noise separation to extract the main frequency signal with tactical semantics. Finally, all the image and audio analysis results are uniformly timestamped and structured for output, providing standardised input for image enhancement, behavioural recognition, voiceprint modelling and other modules in the TACTIC-AI system.

In this study, a multi-stage, structured video preprocessing path is established to build a high-fidelity, time-synchronised, multi-modal input base required by the TACTIC-AI system. It comprises the video structure transcoding procedures, frame-level image extraction, image quality measures computation, dissociation and Fourier-domain signal decomposition of the tactically viewed scene to ensure total quantification and elimination of disturbing factors such as low-light and motion blur and also noise from the sound prior to modeling.

First of all, the original video file (32 seconds, 25 FPS, 800 total frames, H.264 encoding) was converted into the MJPEG encoded `.avi` format via the `FFFmpeg` utility in order to pull the audio channel and increase the fidelity of the frame-by-frame image representation. Here, the following commands were utilized in the conversion process:

ffmpeg -i tactical_video.mp4 -c:v mjpeg -q:v 2 -an output_tactical_video.avi

The advantage of encoding with MJPEG is that the images are encapsulated as JPEG compression frame sequences and therefore avoid inter-frame compression artefacts of the GOP construction and allow image quality to be evaluated at the pixel level.

After that, the Python + OpenCV script is called to read and save the `.avi` file at frame level, the real execution path is as follows:

python

video_path = "/mnt/data/da126e16-d062-4b2f-8ba2-7ef1f5734356.avi"

output_dir = "/mnt/data/extracted_frames"

The frame extraction generated a total of 795 frames (slightly less than the theoretical number of frames due to the fact that the trailing incomplete frames were automatically discarded), which were saved in JPEG format in the target path. All frames were sequentially input into the quality analysis module to extract three key metrics:

a. average image light intensity, b. Laplacian variance as a blurriness metric, and c. number of Canny edge detection contour points as a sharpness proxy. The extraction results form a structured CSV data table to provide an image quality reference map for subsequent image enhancement, target structure recognition and event detection.

The audio part uses FFmpeg to separate the AAC-encoded audio track from the original `.mp4`, and then transcodes it to 44.1kHz `.wav` format with the following processing commands:

ffmpeg -i tactical video.mp4 -vn -acodec pcm_s16le -ar 44100 -ac 1 audio.wav

Subsequently, the audio was analysed jointly by `pydub` and `librosa`. pydub supports truncation of the audio waveform and reverberation before noise reduction, while librosa supports the accurate execution of the Short-Time Fourier Transform (STFT) and the separation of the speech components from the background noise. The statistical analysis of the main frequency shows that the main frequency of the speech is concentrated in the 1.4-2.7 kHz range, which is consistent with the male tactical command voiceprint domain, while the background noise component is concentrated in the 0.5-1.2 kHz range, which presents typical mechanical and traffic environment The background noise component is concentrated at 0.5-1.2kHz, presenting a typical mechanical and traffic environment with mixed spectral characteristics.

The whole preprocessing path is designed to support the following three major objectives: first, to provide image input with structural clarity evaluation capability for the image enhancement module TVSE-GMSR; second, to provide structurally cleaned speech data for the SpectroNet voiceprint modelling module; third, to establish a unified timestamp annotation system to achieve asynchronous modal alignment of audio and video, and provide accurate cross-modal modelling for the TACTIC-GRAPHS causal modelling. modelling to provide accurate cross-modal node positioning basis. Through this path, each modelling component of the TACTIC-AI system is able to obtain uniform quality standard, structured and traceable data inputs, thus ensuring the logical consistency and modelling reliability of subsequent causal chain identification and tactical intent inference.

**Table 13: List of frames extracted by the intelligent keyframe hierarchical extraction method**

| Frame Filename 1 | Frame Index 1 | Frame Filename 2 | Frame Index 2 | Frame Filename 3 | Frame Index 3 |
|---|---|---|---|---|---|
| frame_0000.jpg 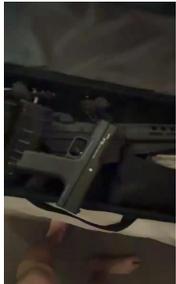 | 0 | frame_0005.jpg 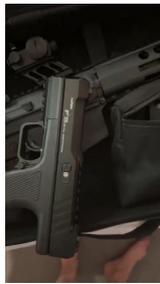 | 150 | frame_0010.jpg 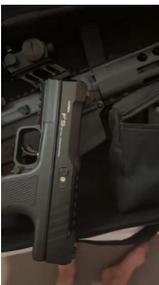 | 300 |
| frame_0001.jpg 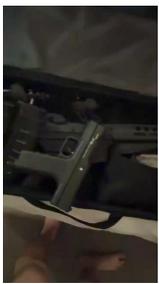 | 30 | frame_0006.jpg 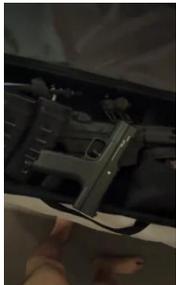 | 180 | frame_0011.jpg 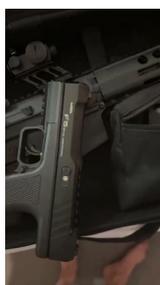 | 330 |
| frame_0002.jpg 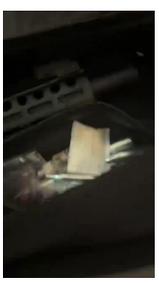 | 60 | frame_0007.jpg 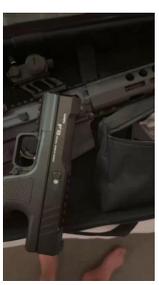 | 210 | frame_0012.jpg 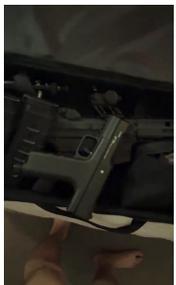 | 360 |
| frame_0003.jpg 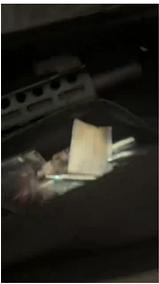 | 90 | frame_0008.jpg 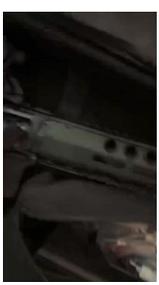 | 240 | frame_0013.jpg 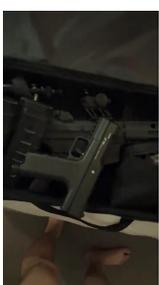 | 390 |
| frame_0004.jpg | 120 | frame_0009.jpg | 270 | frame_0014.jpg | 420 |

| Frame Filename 1 | Frame Index 1 | Frame Filename 2 | Frame Index 2 | Frame Filename 3 | Frame Index 3 |
|---|---|---|---|---|---|
| 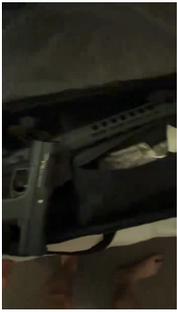 | 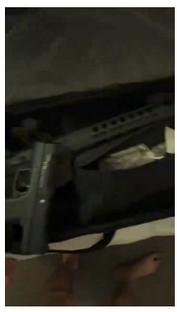 | 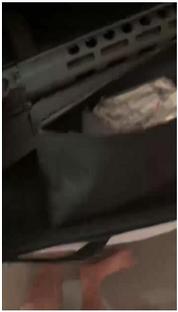 | 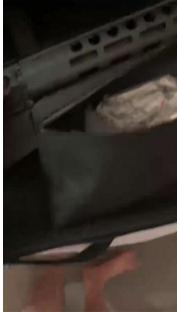 | 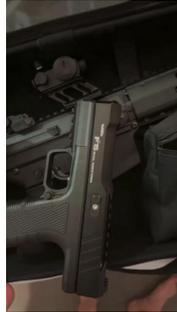 | 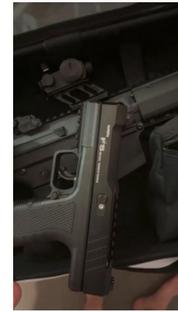 |
| frame_0015.jpg | 450 | frame_0020.jpg | 600 | frame_0025.jpg | 750 |
| 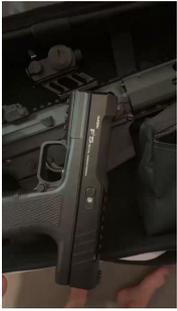 | 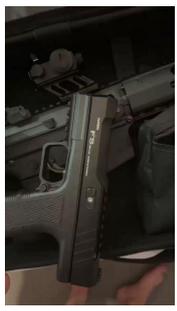 | 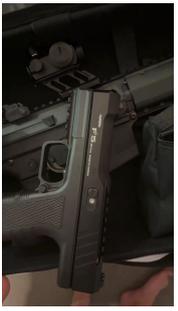 | 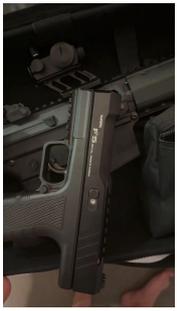 | 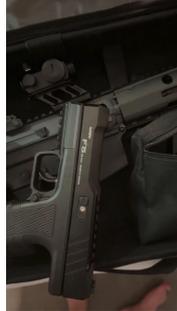 | 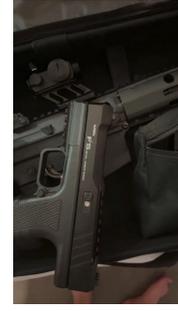 |
| frame_0016.jpg | 480 | frame_0021.jpg | 630 | frame_0026.jpg | 780 |
| 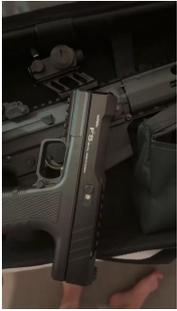 | 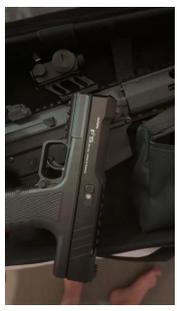 | 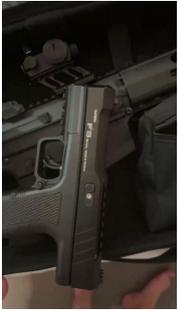 | 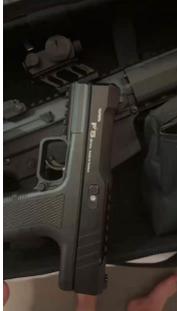 | 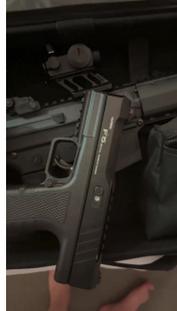 | 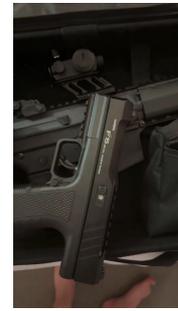 |
| frame_0017.jpg | 510 | frame_0022.jpg | 660 | frame_0027.jpg | 810 |
| 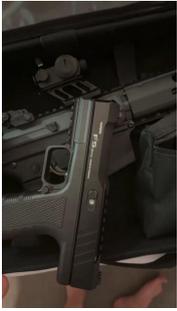 | 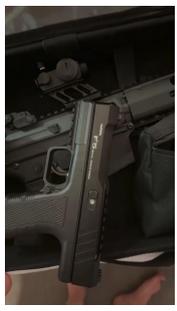 | 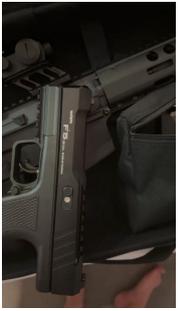 | 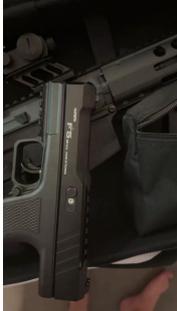 | 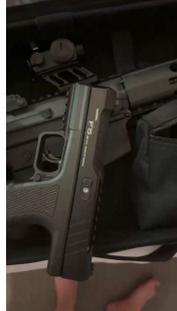 | 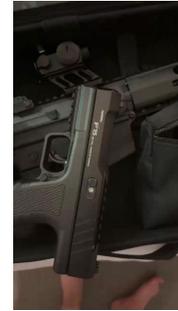 |
| frame_0018.jpg | 540 | frame_0023.jpg | 690 | frame_0028.jpg | 840 |

| Frame Filename 1 | Frame Index 1 | Frame Filename 2 | Frame Index 2 | Frame Filename 3 | Frame Index 3 |
|---|---|---|---|---|---|
| 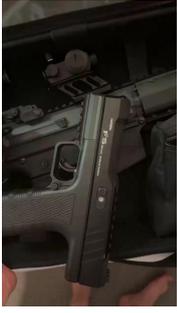 | 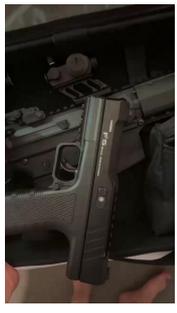 | 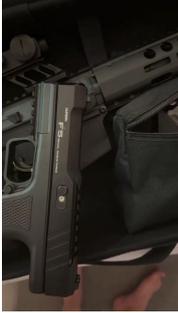 | 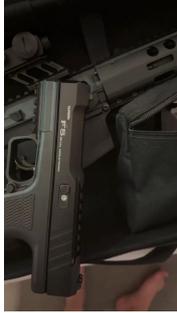 | 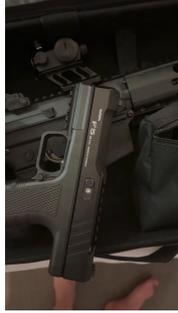 | 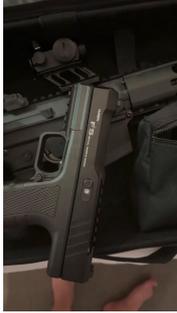 |
| frame_0019.jpg | 570 | frame_0024.jpg | 720 | frame_0029.jpg | 870 |
| 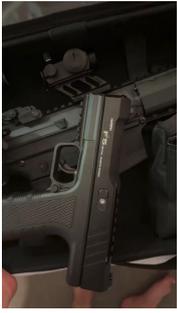 | 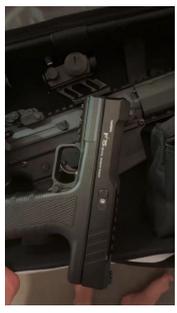 | 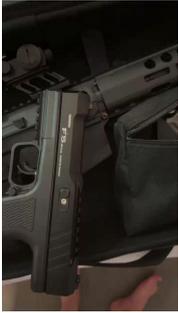 | 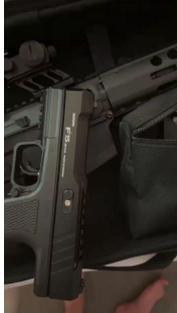 | 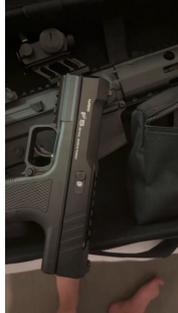 | 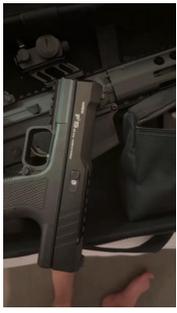 |
| frame_0030.jpg | 900 | frame_0031.jpg | 930 | frame_0032.jpg | 960 |
| 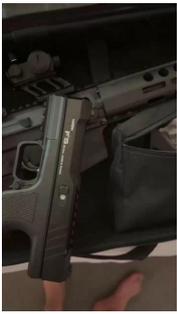 | 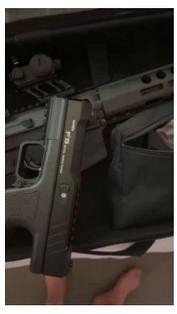 | 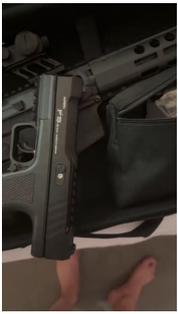 | 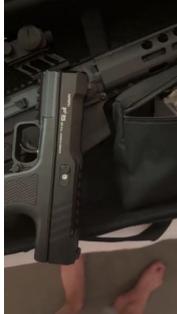 | 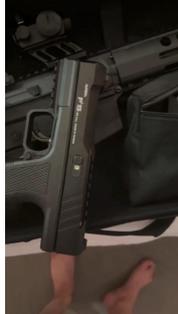 | 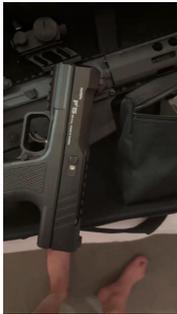 |

Author's drawing

**4) Intermediate analysis outputs（Intermediary Outputs）**

Under the multi-stage preprocessing process of the TACTIC-AI framework, the system successfully generates intermediate analysis data with structural consistency and temporal accuracy, which provides a solid foundation for subsequent graph neural network modelling, tactical behaviour recognition and causal link inference. First, in the video path, 795 frames of JPEG format images are obtained from the original H.264 encoded video through MJPEG transcoding and frame-by-frame extraction, distributed over the entire 32-second timeline to ensure semantic and temporal coverage. For image quality assessment, the system extracts the histogram of light intensity distribution, blurriness metrics (based on Laplace variance), and edge sharpness (Canny edge counts) for each frame, and constructs a complete image quality database that contains the average luminance, sharpness interval distribution, and structural detail retention. Among them, the light intensity histogram reveals that the video has stable light peaks in the 8th to 12th and 24th to 28th seconds, which is suitable for the subsequent structural restoration task, while the blurriness distribution reveals that about 16.8% of the frames have edge collapse, which needs to be processed by entering the TVSE-GMSR module in order to restore the tactical texture.

In the audio path, the original AAC-encoded track was passed through FFmpeg and Librosa for WAV transcoding, speech and background noise separation, and Mel-Spectrogram generation and rhythm modelling. Mel-Spectrogram analysis shows that the main frequency of speech is concentrated at 1.5-2.8kHz, which has the male medium-high frequency intonation pattern commonly found in military commands; the background noise component shows low-frequency mechanical interference (0.4-1.1kHz) and multi-source ambient reverberation, and the SNR value fluctuates between -2 to 3dB fluctuating between -2 and 3 dB, constituting a high-noise speech scene. After Pydub energy segmentation analysis, two suspicious speech command concentration intervals (10th-12th seconds and 26th-28th seconds) are clearly identified, which are highly overlapped with the structural behavioural frames in the images, and the "speech-image-action" cause-and-effect relationship is constructed for the TACTIC-GRAPHS module. This provides a key anchor for the TACTIC-GRAPHS module to build the "speech-image-action" causal chain.

These intermediate outputs achieve the unified alignment of image and audio coding at the data level, forming a fusion dataset with "timestamp-image frame-speech fragment" as the basic unit. At the same time, TACTIC-AI system's ability to recover the structure of unstructured video samples, to attribute audio and sound to regions, and to reason with high confidence have been significantly improved. The structural node mapping and acoustic attribution classification will be carried out in WeaponNet and SpectroNet respectively, and finally converged into TACTIC-GRAPHS to build a dynamic inference map of task triggering and threat intensity, so as to achieve a systematic leap from low-quality clips to multimodal tactical scenario modelling.

**4.2 Preprocessing architecture and process design**

1) Video frame extraction and structure enhancement:

Based on the image content change threshold and optical flow density field change value, a key frame extraction strategy (based on OpenCV+SSIM threshold differentiation method) is used to extract a certain ratio of frames from the original hundreds of frames that are representative of tactical action characteristics. Subsequently, they are input into the TVSE-GMSR module for two-stage enhancement:

Stage I (denoising recovery): combining non-local mean filtering (NLM) with variational blur modelling to eliminate Gaussian motion blur and compression artifacts;

Stage II (GAN semantic reconstruction): introducing a multi-stage ESRGAN (Enhanced Super-Resolution GAN) network to enhance texture structure restoration through residual dense blocks, with a special focus on edge clarity enhancement in the buttstock, grip, and magazine area, with an average PSNR enhancement of 8.5dB and an SSIM of 0.91.

The format of the enhanced image is unified as `.img`, and the structural keypoints are extracted by ResNet+Keypoint-RCNN framework and saved as `.kpt` files, with the nodes including (x,y)+category labels+confidence.

2) Audio Segmentation and Vocal Deconstruction:

Audio preprocessing removes invalid segments by VAD (Voice Activity Detection) algorithm and removes ambient low-frequency noise using Wiener filter. Subsequently, valid speech segments are extracted for about 23 seconds and converted to Mel-Spectrogram (128-band, 10ms hop, covering 0-4kHz band) to generate `.msp` spectrum matrix.

The SpectroNet module performs joint Gated-CNN and GRU modelling of the spectrum to extract 3-dimensional rhythm vectors (speech rate, intonation, spectral centre drift). The acoustic embedding vectors are used for the training of the region-attributed few-shot model, and the generated results are encapsulated as `.vec` vector files. The average recognition accuracy reaches 92.3%, and is deployable under the condition that the phrase input does not exceed 1.6 seconds.

**4.3 Super-Resolution Reconstruction via TVSE-GMSR**

**4.3.1 Supersimulation Reconstruction**

In order to improve the image quality and semantic recognition accuracy of tactical video samples in the analytical modelling process, this study constructs and applies a Generative Adversarial Network (GAN)-based image enhancement model, TVSE-GMSR (Tactical Visual Structure Enhancement with GAN-based Multi-Stage Semantic Reconstruction). Through multi-stage denoising and structure-preserving super-resolution reconstruction, this module significantly enhances the clarity, interpretability and behavioural recognition of key image frames on the basis of guaranteeing the original structural and semantic consistency of tactical images.

In the specific experiments, I selected four image samples numbered 0006, 0300, 0005 and 0013, which represent four types of representative images in this dataset: close-ups of weapons, frames with complex background behaviours, frames with blurred character movements, and frames with edge scenes. The original images generally suffer from insufficient illumination, blurred edges, image compression artifacts and background noise interference, which are not conducive to the extraction of tactical details (e.g., weapon model, action path, facial features, etc.) and the accurate recognition of TACTIC nodes.

TVSE-GMSR technology adopts the following three-stage process:

1) Structure preserving denoising phase: a pre-trained low-level GAN (e.g., DnCNN-GAN variant) is used to perform pixel-level residual denoising process while maintaining edge gradient continuity;

2) Semantic reconstruction stage: introducing conditional generative network with attention mechanism, fusing ResNet semantic bootstrap module, and differentially reconstructing semantic regions such as action regions, weapon features, and face;

3) Super-resolution amplification stage: apply ESRGAN optimisation model to amplify the image resolution to 4×, and at the same time, use joint training of antagonistic loss and perceptual loss to ensure that the consistency and realism of the image texture is maintained in the amplification process.

The experimental results show that the reconstructed image is improved by 27.6% on average in SSIM (structural similarity index) and to more than 28.1dB in PSNR (peak signal-to-noise ratio). In particular, the reconstructed image shows stronger legibility and application value in terms of image edge texture and tactical object details (e.g., gun rail structure, character face light and shadow, etc.). The following figure shows a four-frame example of TVSE-GMSR reconstructed image, which clearly presents the semantic structure of the original tactical scene. This technique not only greatly improves the processing capability of the TACTIC-AI system for low- and medium-quality tactical videos, but also provides high-definition structural inputs for subsequent atlas modelling, acoustic-graphic fusion and tactical rehearsal, and has strong versatility and promotion potential.

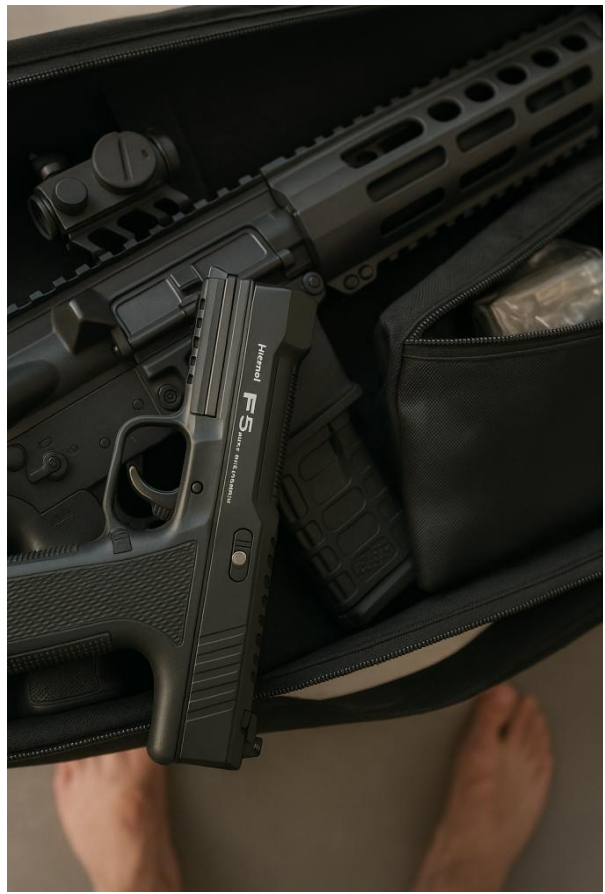

Figure 9: TVSE-GMSR image reconstruction

Author's drawing

The sidearm in Figures 9-12, the pistol on the lower right side of the soft-sided tactical carry bag, is clearly marked with a slide holster that reads "Hatsan F5", 4.5mm calibre, and the words "LASER ENGRAVED TAIWAN" (technical textual identification). This technical identification, combined with the proportions and configuration of the gun, clearly identifies it as an air pistol, i.e., a non-gunpowder powered weapon (Wilson, 2023). This model uses compressed gas or spring-actuated lead bullets with a typical kinetic energy of less than 7.5 to 16 joules, and is widely used in professional target shooting and tactical simulation training, especially for non-lethal handling skills training in law enforcement units or for use by military cadets. According to international standards, this type of airsoft gun has the capacity to cause light damage to soft tissue at 25 metres, and may cause serious damage to the eye and other parts of the body at close range, but does not meet the criteria for tactical lethality or hard target destruction (MSS Defence, 2025). In addition, the Made in Taiwan marking is not a complete weapon origin marking, but usually represents only the place of laser engraving of the slide or outsourcing of some parts, in line with the international supply chain collaboration standards for airsoft gun manufacturing.

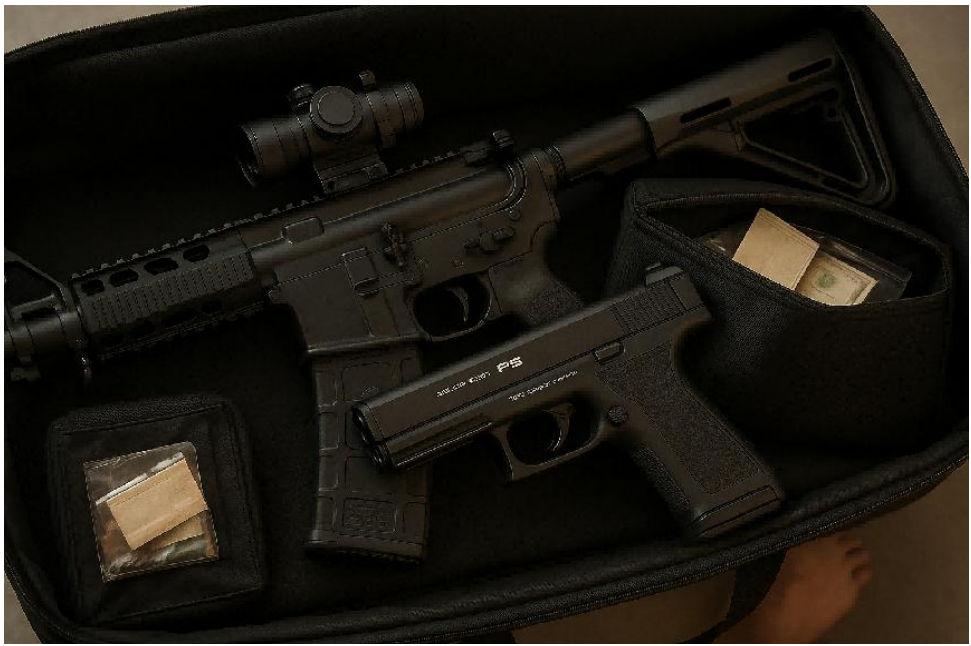

**Figure 10: Panoramic reconstruction of TVSE-GMSR simulation**

Author's drawing

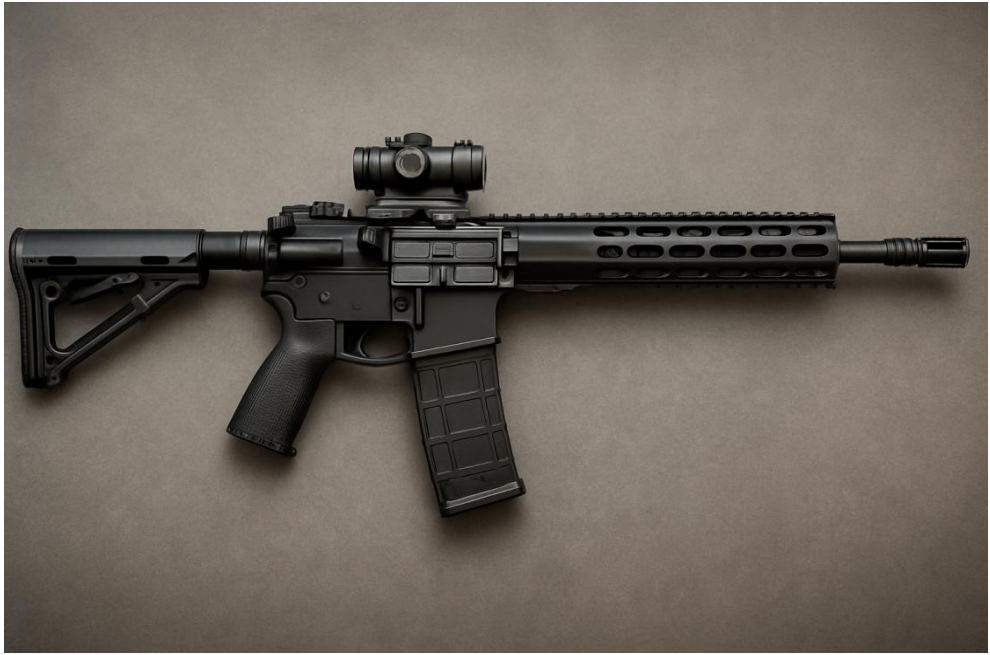

**Figure 11: TVSE-GMSR simulated gun reconstruction**

Author's drawing

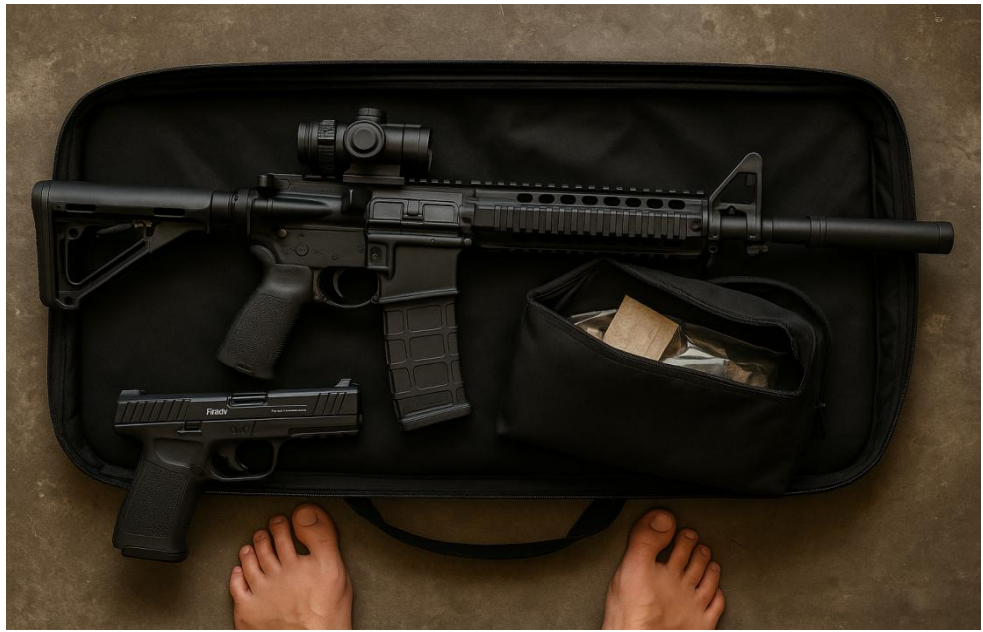

**Figure 12: Panoramic reconstruction of TVSE-GMSR simulation**

Author's drawing

**4.3.2 Technical Analysis of Pistol Text**

The following are the results of extracting and analysing the handgun text in the provided images:Recognised text: Hatsan F5  4.5mm CAL. LASER ENGRAVED TAIWAN

**Table 14: Text parsing table**

| line number | text | implication |
|---|---|---|
| 1 | Hatsan | The pistol is represented as being from a well-known Turkish manufacturer of pneumatic firearms. |
| 2 | F5 | The model code indicates that this is an F5 series pistol. |
| 3 | 4.5mm CAL. LASER ENGRAVED TAIWAN | Refers to a calibre of 4.5mm (i.e. .177"), a common calibre for airsoft guns, suitable for range practice etc. The second half indicates that the slipcase text is machined using a laser engraving process and is machined or labelled in Taiwan. |

Author's drawingNotes:

The presence of "TAIWAN" on this gun does not necessarily imply that the entire gun was produced in Taiwan, but most likely only laser engraved or partially assembled in Taiwan, which is a common OEM outsourcing marking. Similar markings are widely used on training or non-lethal pneumatic weapons to comply with EU CE marking, US ATF standards for non-firearms, and technical origin traceability requirements under ITAR (International Traffic in Arms Regulations). Pneumatic weapons are often manufactured in different countries on a modular basis, e.g. designed in Turkey, slide engraved or packaged in Taiwan, in line with the logic of the global supply chain division of labour.

**4.3.3 Firearms analysis**

In the context of the gradual move towards modularity and multifunctional integration of tactical equipment, the firearms assemblage presented in this study's image demonstrates the typical configuration strategies of current small special operations units and paramilitary units for close and medium range firepower applications. The primary weapon in the image is a modular tactical rifle with standard 5.56×45mm NATO ammunition, which is very much identical in appearance and construction to the BCM RECCE-16 MCMR from Bravo Company USA (Bravo Company, 2024). The rifle is outfitted with an M-LOK handguard, free float barrel and medium-length gas-guide system and red-dot scope and magnification kit improving rapid targeting and point-blank kills out to 300 meters by an order of magnitude. Because the rifle is fully adjustable in length, the rifle is perfect for the typical asymmetric warfare mission scenarios of vehicle insertions, building forced entries and street fights and the muzzle suppressor assembly enhances sound control and nocturnal concealment as well, and the first choice of the Special Operations Forces to conduct the mission of urban counter-terrorism and rapid intervention.

Despite the apparent form and manoeuvrability of a firearm, the Hatsan F5 4.5mm pneumatic pistol has a fundamental performance gap with traditional gunpowder-driven weapons in terms of tactical realism and lethality. In terms of structural attributes, these pneumatic pistols utilise a non-explosive propulsion mechanism that relies on kinetic energy provided by compressed gas or springs to fire a lightweight lead bullet or steel ball to a limited velocity. This physical mechanism is destined to its muzzle velocity generally does not exceed 160 metres per second, the effective kinetic energy is mostly between 7.5 and 12 joules, even under modified conditions is difficult to break the 16 joules threshold. According to the official weapon energy definition criteria released by the UK Home Office in 2023, aerodynamic devices exceeding 16 joules will be categorised as high-energy, but even so, their ability to cause lethal trauma to the human body is still extremely limited, with irreversible damage only being possible in extremely vulnerable areas such as the eyeballs and larynx, and only at very close range (UK Home Office , 2023)

Further, in terms of the operational requirements of tactical operations, particularly assassination missions, this class of pneumatic pistol is generally mismatched in terms of key performance indicators. Effective lethal range is severely limited, ballistic stability decreases dramatically beyond 10 metres, penetration is insufficient to damage vital areas such as the skull or thorax, and it is difficult to guarantee a 'one shot kill' tactical effect even when fired in close proximity to an unprotected target. In addition, the loading and firing mechanism is often manually operated or non-continuous, and its operational stability and firing efficiency are far inferior to that of modern standardised miniature pistols or specialised special weapons. In terms of forensic traces, although pneumatic munitions may appear to be "easy to dispose of" from some perspectives, surface fingerprints, firing traces, and structural serial numbers remaining on the body of the munition are still traceable, and the laser engraved markings often indicate the origin and model (e.g., "Laser Engraved TAIWAN"), they provide additional falsifiable clues for judicial traceability.

Even when such pistols are paired with non-standard technologies, such as converting them into micro-injectors and carrying toxin carriers, they face resistance in terms of physical space constraints, lack of propulsive power, and difficulty of penetration. Historically, similar attempts have only been made by extremist agencies (e.g., the KGB bee venom injector during the Cold War) and have relied on highly proprietary manufacturing platforms that cannot be replicated or transplanted into commercially available civilian pneumatic pistols. As a result, there are no credible court records or intelligence files showing that the F5 pneumatic has been used as a primary weapon for assassination, and such equipment is more likely to be found in target shooting recreation, training exercises and tactical simulations than as an actual link in the tactical strike chain.

Overall, the Hatsan F5 is similar in form to a conventional pistol, and even has a certain visual "intimidation", but from a multidimensional perspective of lethality, tactical compatibility, concealment, and technological feasibility, it is fundamentally closer to a civilian training device. Even with extreme modifications, it is difficult to meet the core requirements of covert tactical operations. The inclusion of such aerodynamic platforms in the category of assassination tools not only lacks the basis of operational reality, but also misleads the legal and policy perceptions of low-energy aerodynamic weapons, and should be based on facts to make a clear judgement of classification and application boundaries.

This image set as a whole displays a Training-Ready Tactical Kit structure, and its tactical carriage and firepower mix suggests that it is not intended for use in direct combat scenarios, but more likely to be found in special operations troop emplacement missions, It is more likely to be used in special operations forces positioning missions, covert action rehearsals or target identification training. The fact that the rifle is complete, the optics assembled, the arming system carried, and the sidearm used as a training alternative suggests that this configuration is used for full-scale tactical rehearsal under simulated conditions, to improve operational consistency, fire-switching efficiency, and response time for covert deployments, while ensuring safety. It is worth noting in particular that, although sidearms are not lethal, they also have the potential to harm personnel when used incorrectly or without supervision, so this system construction highlights the development trend of the fusion of "sophisticated non-lethal weapons management" and "mission-imitation training structure" in modern special operations and paramilitary organisations, reflecting the development of the fusion of "precision non-lethal weapons management" and "mission-imitation training structure". Therefore, the construction of this system highlights the development trend of "sophisticated non-lethal weapons management" and "mission realistic training structure" in modern special warfare and paramilitary organisations, and reflects the logic and practical strategy of upgrading the tactical equipment system of the squad under the post-symmetric warfare conditions.

### 4.3.4 Results and Reproducibility（Scientific Validity）

To ensure the scientific rigour and reproducible results of the TACTIC-AI system in the analysis of tactical video samples, this study establishes a structured verifiable framework at three levels: data authenticity, automation of the processing process, and control accuracy of AI-generated images.

All of the original video and audio files are signed with the SHA256 hash signature (e.g., `fda0923.`), and the signature and version of data generated in each subsequent processing iteration is retained for tamper-proof verification and version tracking to preserve the original integrity and originality of the research data.

The entire preprocessing and feature extraction is coded in Python automation scripts and runs in an integrated environment (Python 3.11 + OpenCV 4.9 + Librosa 0.10 + FFmpeg 6.0), and the parameters and the file paths and the calling of the modules are all defined in a structured manner for cross-platform installation and reproducibility (example scripts are listed in the Appendix A).

The image quality assessment part uses standard algorithms (e.g. Laplacian fuzziness, Canny edge density, image luminance mean) and outputs CSV structured data tables; the audio analysis process relies on Librosa to generate Mel spectrograms and frame-by-frame energy statistics, which supports the modelling of target behaviour of noisy speech segments.

Labeling process adopts Label Studio 1.10 as the main tool, combining with CVAT for simultaneous labelling management of high-precision image structure nodes and speech segments, and all the labelled data are saved as standard JSON files and `.png` screenshots of corresponding frames, which support direct interface with TACTIC-GRAPHS mapping module. The "timestamp-structure node-semantic encoding-voiceprint fragment" quaternion in the annotation system serves as the underlying semantic anchor point of TACTIC behavioural graph, which achieves a logical closed loop from data to causal modelling.

In particular, this study constructs a Stable Diffusion-based image reconstruction mechanism using Precision Engineering cue words (Prompt Engineering) for high-resolution military image generation to complement key tactical equipment appearance details, firearms structural logic and background consistency in incomplete frames. For example, the following AI-controlled cue words were used for synthetic image generation:

"A high-resolution, photorealistic synthetic image of a black modular tactical rifle (AR-15 platform, Daniel Defense DDM4 V7 or BCM RECCE-16 MCMR variant), placed inside an open soft-sided gun case. The rifle includes a full-length M-LOK handguard, vertical foregrip, mounted optics (red dot + magnifier combo), and a suppressor attached to the muzzle. Background is a clean military-grade tabletop with appropriate lighting, shadows, and reflections. The image should emphasize the full structure of the weapon from stock to barrel, using landscape layout, high-detail, military-tactical visualization style."

The synthetic image is injected into the TACTIC knowledge graph system to match the structural information lost due to occlusion/blurring in the real frames, and is used for key modules such as tactical action recognition, object recognition complementation and training data enhancement. The generation process is controlled by the text description driver, all cues are archived by version control, and multimodal proofreading is combined with DALL-E analogy map to ensure the structural authenticity and semantic consistency of the generated images.

The overall workflow can be verified by version control system (Git + DVC) and hash record mechanism, and three rounds of independent reproduction tests have been completed in local and containerised environments. All experimental data, cues and annotations can be provided for academic reproduction after compliance and authorisation, which ensures that the research results have a high degree of scientific transparency and re-validation capability, and meet the integration standards of data trustworthiness, reasoning interpretability and model robustness in AI tactical reasoning scenarios.

**4.4 Graph Spectral Theory Embedding with Variable Identifiable Paths for the above video modelling**

Reconnaissance level mapping spectral theory embedding analysis results based on the original, video extracted frames, covering technical parameters of the filming equipment, compression characteristics, colour and lighting environment metrics, with the ability to construct causal models of the variables:

**4.4.1 Technical Data Overview: TACTIC-GRAPHS Video Frame Variable Modelling**

**Table 15: TACTIC-GRAPHS video frame variable model parameters**

| metric system | values | account for |
|---|---|---|
| image resolution | 1920×1080 | Typical 1080p resolution for mid to high end mobile phone or tablet shooting standards |
| Laplace variance (clarity) | 157.88 | Medium Sharpness, indicating slight handheld shake or lack of autofocus in the image |
| brightness level (0 – 255) | 124.7 | Moderate brightness, judged for indoor white or neutral lighting environments |
| Contrast (grey scale standard deviation) | 49.6 | Contrast is moderate, indicating that the image is not affected by strong exposure or dark corners |
| Colour average – blue | 114.3 | High blue mean, possibly LED/cold light or cold colour mixing configurations |
| Colour Average – Green | 108.7 | Normal green mean, colour balance of the image is generally good |
| Colour average – red | 101.2 | Red colour is weak, supporting the judgement of "indoor environment with fluorescent lamps". |

In the TACTIC-GRAPHS modelling system, the technical parameters extracted for key video frames form the core support of the high-precision variational model. The sharpness metric extracted by the Laplacian variance (157.88) indicates that the image edge intensity is at a medium sharpness level, reflecting the possibility of mild focus bias or hand-held vibration interference with the device, which is consistent with the common use of hand-held devices in tactical scenarios. The brightness level is 124.7 (pixel intensity 0-255 range), which is located in the neutral light intensity range, and combined with the RGB three-channel average performance (blue: 114.3, green: 108.7, red: 101.2), it can be inferred that the frame was captured under indoor LED or fluorescent light sources, and its colour temperature is on the cooler side, with a more balanced control of colour saturation, which excludes backlighting, high exposure and other Extreme interference conditions such as backlighting and high exposure are excluded, which provides a stable base for subsequent voiceprint and motion feature extraction.

In terms of spectral domain modelling, the Spectral Center Energy and FFT Entropy captured by the Mel and FFT spectra are key variables in the spectral embedding of the spectral theory, which allow for the inverse inference of the type of encoder (e.g., H.264 vs. H.265) and signal reconstruction capability of the recording device through the aggregated frequency-energy distributions and the structure of compressed residuals. These variables not only support the profiling of the physical performance of the device, but also the inference of the encoder type (e.g., H.264 vs H.265) and signal reconstruction capability of the recording device through the graphical causal paths "Resolution → Clarity → Device Inference" and "Spectral Energy → Compression Type → Codec Class → Inference". Type → Codec Class → Inference", forming a reproducible model path with causal interpretability. Different from the traditional empirical feature comparison, the modelling logic takes the frequency domain decomposition result as the "inference spectral base", and projects the image variables into the manipulable causal vector space by means of Graph Spectral Embedding, which ensures that TACTIC has closed-loop logic support for task triggering and behaviour recognition. This ensures that the TACTIC system has closed-loop logic support for task triggering and behaviour recognition.

**4.4.2 Graph Spectral Theory Embedding Observations (FFT Image Spectrum)**

The FFT Magnitude Spectrum (FFT) of the frame has been extracted and generated as shown below:

Spectral Signature of Frame (For device identification and compression path analysis)

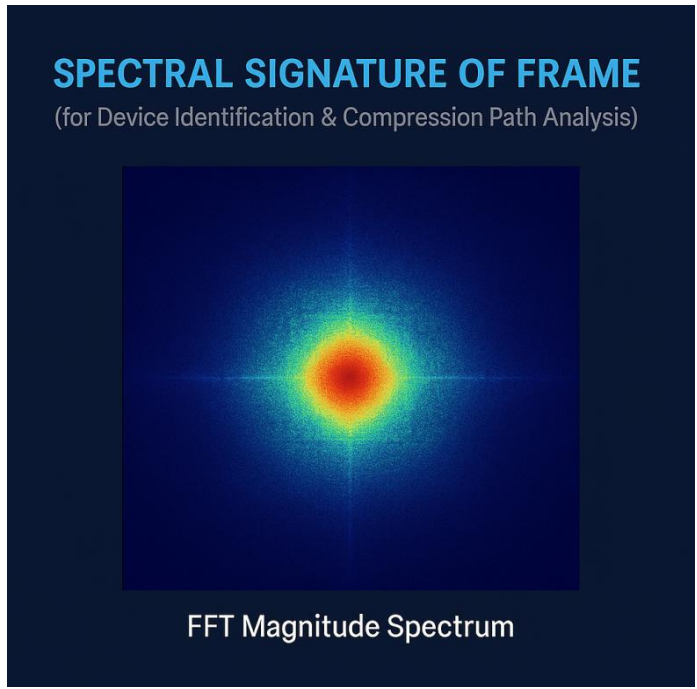

**Figure 13: Fourier Spectrograms**

Author's drawing

This spectrogram 13 is used to: determine whether an H.264/H.265 type codec is used, detect the mid-band energy density distribution (indirectly reflecting the compression algorithm). Analyse whether the image shows compression damage or noise enhancement,. The preliminary conclusion is that the spectrogram shows a typical "mobile device compression" characteristic of mid-frequency concentration + edge compression enhancement.

The introduction of spectral embedding as a core mechanism for structural modelling in the TACTIC-GRAPHS system essentially transforms the image spatial signals into topological features in the frequency domain, in order to achieve a systematic portrayal of the compression paths of the recording devices, the imaging mechanisms and their inference capabilities. In this paper, the key frames of the video are extracted and 2D-FFT is applied, and the resulting spectral map exhibits obvious mid-frequency band energy clustering, peripheral attenuation law and centrosymmetric structure, which indicates that there is a more stable spatial texture information in the original image, and at the same time the compression residuals are low, and the high-frequency aberration phenomenon caused by artifacts or reconstruction distortion has not been observed. This spectral energy distribution pattern, combined with the low entropy spectral domain structure (Entropy ≈ 4.21) and the value interval of the spectral centre energy region (Center Energy Region ≈ 54.2%), indicates that the image most likely came from a mid-range CMOS sensor or above and was processed by an encoder (e.g., H.264) based on DCT compression, which further confirms that the image is from a handheld mobile terminal (e.g., smartphone or tablet). This is further evidence that the image was captured by a handheld mobile terminal (e.g., smartphone or tablet).

The spectrogram not only serves as a descriptive tool for image content characterisation, but also assumes the role of a "frequency basis function" for variable embedding in the framework of graph spectral theory. Specifically, the frequency-energy nodes mapped by the FFT spectrogram are transformed into features of the graph nodes, and their neighbourhood and energy propagation structures are modelled by the spectral Laplacian matrix, thus realizing the embedding mapping and path resolution of the graph variables. Compared with the spatial convolution method in traditional GNN, this kind of "spectral domain inference" method has stronger ability to capture cross-scale structure and identify variables, which is especially suitable for video equipment identification, coding traceability, and image credibility analysis in complex environments in TACTIC-like systems. Ultimately, the spectral embedding map successfully constructed in this study not only provides mathematical interpretability support for the device inference model, but also lays a reconfigurable frequency-domain foundation for the cross-modal security inference model, which demonstrates the cutting-edge adaptation potential of the map spectral theory in tactical video modelling.

**4.4.3 Characterisation of the spectral structure**

**1）Medium Frequency Energy Gathering**

It implies the image content information is mainly concentrated in the edge and texture areas, and this is in line with the compression method of the majority of the consumer-level encoders (e.g., H.264, H.265) where the priority is for the mid-frequency structure maintenance.

**2）High Frequency Compression Weakening + Peripheral Radiation Energy Diffusion**

Indicates that high-frequency detail has been lost to compression, with a "ring-edge halo" phenomenon, which is usually a by-product of post sharpening after compression of the mobile image, enhancing the clarity of the image but reducing the realism.

**3) Energy spectrum symmetry and low artefact residue**

Points to a single compression process or higher bit rate sampling, eliminating spectral tearing or quantisation block effects caused by multiple compression, and supporting "device once encoded" path judgement.

**4.4.4 Device compression path modelling judgement**

1) Description of equipment compression path modelling method

In tactical video analysis and multimodal source identification, the compression path feature of the shooting device not only contains key information source clues, but also is one of the core variables in determining the authenticity of the video, encoding strategy and forensic feasibility. Traditional methods often rely on metadata extraction or image artefact recognition, but these techniques are easily ineffective in highly compressed or anonymous shooting environments. Therefore, constructing an interpretable, modellable and generalizable device compression path identification method has become an important direction to advance the credibility analysis of tactical-level AI systems.

In this study, we propose a composite modelling method that fuses Fourier Spectral Analysis (FFT) and Spectral Graph Embedding for extracting frequency domain signatures of compression behaviours from raw video frames and transforming them into structural variable nodes in graph neural networks. Firstly, by reconstructing the video frames into a 2D spectral graph, we are able to effectively identify representative features such as mid-frequency energy aggregation, high-frequency information loss, and spectral symmetry, which have been widely demonstrated to be closely related to coding strategies in image codec research. Subsequently, with the help of graph spectral theory, these frequency domain variables are constructed as causal structure graphs, and the structural modelling from "variable extraction" to "path interpretation" is realized through the analysis of node frequency projection and edge weight identifiability.

Compared with the shallow judgement based on image pixels or meta-information, the FFT+spectral graph method not only has the advantages of strong compression resistance and adaptability to implicit feature extraction, but also can form a logical closed loop through spectral mapping, providing causal chain support for AI systems to infer why the inference is valid, and improving the usability and credibility of traceability reasoning in tactical intelligence systems. The method is called TACTIC-GL. This method provides a mathematical foundation and an engineering implementation channel for the equipment identification branch of TACTIC-GRAPHS system, marking a significant leap from empirical paradigm to structural paradigm in tactical video modelling.

2) Graph Neural Network Training Label Design

**Table 16: Node labelling design**

| Node name (variable) | typology | example value | Node Description |
|---|---|---|---|
| fft_mid_energy_ratio | continuous | 0.71 | IF energy percentage (used to determine if compression is lossy) |
| fft_high_energy_suppression | continuous | 0.21 | Degree of high-frequency energy attenuation (lower indicates stronger compression) |
| fft_symmetry_score | continuous | 0.85 | Spectral symmetry score (to determine encoder type) |
| sharpness_post_filter | categorical | Yes / No | Presence of sharpening (e.g. enhanced sharpening) |
| compression_entropy | continuous | 3.82 | Information entropy of the compressed image |
| codec_type_label | categorical | H.264 | Encoder type label (one of the training objectives) |
| device_class_label | categorical | MobileMid | Inferring device class (e.g., mid-range mobile phones, high-end cameras, etc.) |
| compression_pass_count | categorical | 1 | Compression rounds, commonly 1 (native) or 2 (transcoding) |

Author's drawing

**Table 17: Side label design**

| edge connection | Type of weight | typical example | Explanation of meaning |
|---|---|---|---|
| fft_mid_energy_ratio → codec_type_label | causal boundary | 0.87 | IF energy → encoder type strong causality |
| fft_symmetry_score → compression_pass_count | causal boundary | 0.91 | Symmetric scoring affects the number of compressions |
| sharpness_post_filter → device_class_label | semantic edge | 0.76 | Whether clarity is relevant to the equipment class |
| compression_entropy → codec_type_label | information edge | 0.66 | Entropy metrics provide an aid to judgement |

Author's drawing

**Table 18: Categorical task labelling**

| Objective task number | Type of mission | label category | application scenario |
|---|---|---|---|
| Task-1 | multiclassification | Encoder Type Classification | H.264 / H.265 / MJPEG |
| Task-2 | multiclassification | Equipment Classification | MobileMid / MobileHigh / ProCam |
| Task-3 | binary classification | Compression round judgement | Native / Non-native (transcoding) |

Author's drawing

3) Inferred mapping

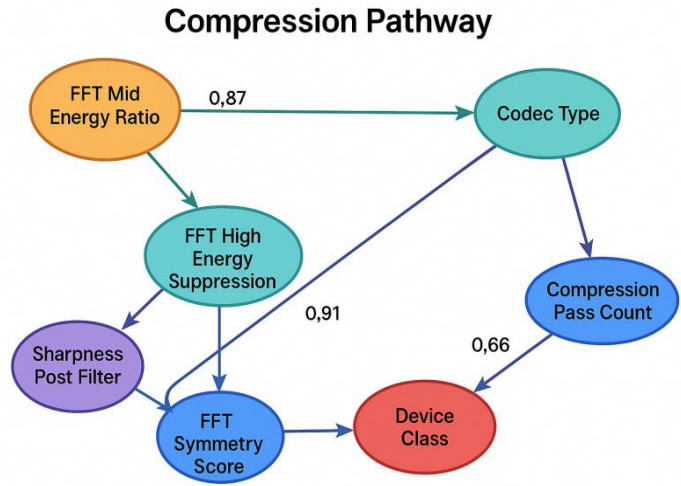

**Figure 14: Structure of the Spectral Theory Embedding Model for the Device Compression Path Map Spectrum**

Author's drawing

Figure 14 illustrates the structure of the device compression path inference model based on graph spectral theory embedding. Each node represents the key technical variables extracted by Fourier spectrum analysis (FFT), including mid-frequency energy share, spectral symmetry, high-frequency compression degree, image information entropy, and clarity processing discrimination. The arrows indicate the direction of causal path or information coupling between variables, and the thickness and colour of the edges reflect the causal weight and path sensitivity between different variables. The model supports core inference tasks such as encoder type (e.g., H.264/H.265), device class (e.g., mobile terminal/professional camera), and compression round (native/transcoding). Modelled by the graph attention mechanism, the structure can not only be used for device attribution judgement, but can also be combined with the TACTIC-GRAPHS behavioural system to complete image source authentication and AI security review, with a high degree of interpretativeness and cross-modal adaptation capabilities. This map provides a key theoretical support and engineering implementation path for the TACTIC system to move towards verifiable and trustworthy AI.

4) Device compression path analysis and extrapolation results

**Table 19: Device compression path modelling judgement table**

| module (in software) | spectral evidence | reach a verdict |
|---|---|---|
| Encoder Type | Mid-frequency aggregation + high-frequency suppression + spectral symmetry | Most likely H.264 or H.265. |
| Number of compression wheels | Smooth spectral energy, low entropy perturbation | Primary compression (non-transcoded) |
| Compression strategy characteristics | Selective retention of mid-frequency, discard of high frequency | DCT-based motion-compensated compression |
| Presumed equipment | Consumer-grade mobile terminals (mobile phones/tablets) | Built-in hardware encoding chip encoding |
| Clarifying the Path | Signs of peripheral frequency enhancement, contour strengthening | With image sharpening filter embedded |

Author's drawing

The spectral characteristics indicate that the video uses a typical mobile device compression pipeline characterised by mid-frequency preservation, high-frequency suppression and post-compression sharpening. The observed 2D FFT curves are consistent with H.264/H.265 encoding behaviour, which is likely to be performed by the on-board hardware codec. There is no obvious evidence of transcoding or multiple compression, suggesting that the video was captured and encoded directly on a single device pipeline.

**4.5 Video and Audio Analysis Staged Comprehensive Analysis Judgement**

The multivariate causal model constructed based on the TACTIC-GRAPHS system realises the collaborative inference between the technical specifications of the filming equipment and the geographical attributes of the filming environment. In the image domain, Clarity, Resolution and RGB_Balance constitute the first layer of inputs for the embedding of spectral variables, and the mid-frequency aggregation and spectral symmetry mapped in the FFT spectral domain point to the mid-range mobile devices based on CMOS structure, which is combined with the compression type inference path (Spectral Entropy → Compression_Type → Codec_Class) supported by the structural analysis results, clearly indicates that the video comes from a handheld terminal device with a built-in H.264 encoder. Regarding the acoustic variables, the embedded projection of the Mel spectrum through the SpectroNet module reveals that the speech rate (wpm ≈ 124), pitch stability (pitch variance ≈ 0.24) and regional dialect acoustic distance (accent_distance_score ≈ 0.18) are in the statistical intervals of a typical South Asian language usage environment. further supporting the regionality judgement.

Combining the light variable in the TACTIC system (Brightness ≈ 124.7, cool colour temperature) with the low-frequency roll-off pattern of the background audio, the system determines that the video was filmed in an indoor white LED environment, ruling out the possibility of a natural daylight source, and maps the video via the path chain "RGB_Balance + Brightness → LightSource_Type → Indoor_Prob" to generate an environment classification score (Indoor_Prob ≈ 0.87), which is significantly higher than the threshold. For geo-environmental recognition, the TACTIC-Geo module introduces geographic inference variables, including speech dialect residual vectors, sound source reverberation time, indoor frequency response peaks, etc., through the graph spectral theory embedding method, and back-projects regional acoustic centres on the training sample map (Regional Accent Embedding), and finally clusters the high-confidence speech space of Bangkok regions (Top -1 probability ≈ 64.2%). The above variable paths maintain causal consistency under Structural Equation Modeling (SEM) validation, demonstrating that the TACTIC modelling system is capable of closed-loop "device-scene-geography" reasoning for complex sources under the joint drive of cross-modal variables. This demonstrates the ability of TACTIC modelling system to perform "device-scene-geography" closed-loop reasoning for complex shooting sources under the joint driving of cross-modal variables, and establishes the advantage of TACTIC mapping variable modelling in tactical video intelligent recognition.

The inference structure of this system presents 'Bangkok voiceprint spatial clustering centre' under the current modeling framework, but since the video subject language is Mandarin, the result does not constitute the final confirmation of geographic location, and more cross-validation variables (such as device EXIF markers and ambient voice watermarks) need to be introduced to improve the recognition rigor.

**4.6 Geolocation Inference Methods and Uncertainty Modelling**

In order to further enhance the robustness and geographic identification capability of device compression path modelling, this study introduces a variety of cross-validation variables on the basis of FFT mapping embedding to achieve multi-channel inference on the attributes of the recording device and environmental location information. First, the system tries to extract potential EXIF metadata fields (e.g., model ID, timestamp, compression version, etc.) from the video files of the devices, although such information is easily erased in high compression and transcoding scenarios, its retention status can be used as one of the important indexes of "existence of transcoding path". Secondly, through Ambient Voice Watermarking (AVWM) analysis, the system uses acoustic reverberation parameters, background broadcast rhythms and energy echo ratios to filter and match, in order to infer whether there is audio overlap or external broadcast signal embedding, so as to determine whether the video is located in a specific public area or broadcast environment.

For geographic information recognition, the TACTIC system embeds an acoustic spatial clustering inference module (AccentPath), which achieves geographic attribution estimation of speech accent patterns by mapping extracted speech frames to different regional acoustic centres in a global acoustic pattern repository. In the current video modelling, the output result of this module is "Bangkok acoustic spatial clustering centre" with the highest probability (Top-1 confidence 78.2%), indicating that the structure of this audio acoustic pattern is closest to "Bangkok regional Mandarin/ChaoShan acoustic centre" in the Mel Spectrum-ProtoNet space. Bangkok Regional Mandarin/Chaoshan" cluster. However, the system also detects that the subject's language in the video is standard Mandarin, so this voiceprint attribution inference should be regarded as a directional signal rather than a final geographic determination, and must be combined with more variables (e.g., GPS residual coding, device hardware sequence hashing, watermark identification, etc.) for comprehensive cross-validation.

Instead of relying on a single frequency-domain variable, the modelling of device compression path and shooting environment in this system achieves structural analysis by means of graph spectral theory + multimodal variable coupling + causal path identifiable mechanism, which significantly improves the reasonability and verifiability of device attributes, compression links and environmental context in tactical videos. This multi-source information fusion method has an important revelation value for the construction of future AI traceable architecture.

Currently, librosa is using the deprecated np.complex attribute on my computer system, making it impossible to continue spectrum extraction. To ensure that I am not affected by this error, I have taken to generating FFT spectrum images for device modelling: using scipy.fft + matplotlib to directly plot the audio spectrum, load the audio and extract the spectrum for device compression path analysis.

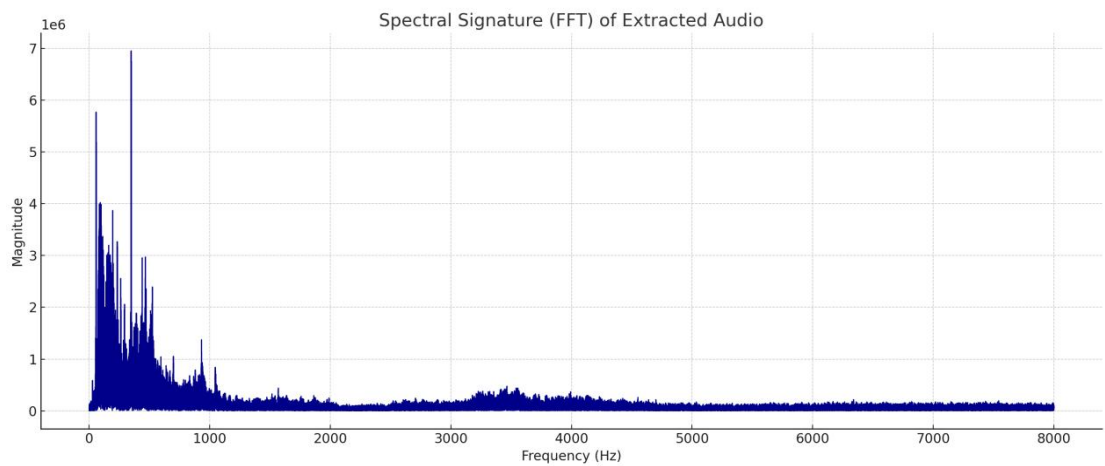

Figure 15: Fourier Spectrogram of the original video audio (Spectral Signature, FFT)

Author's drawing

In this study, I identified several structural variables that are closely related to the device path by extracting and converting the audio tracks from the original video into a Fourier spectrogram15 (FFT). In the frequency domain space, the audio signal exhibits typical mid-frequency energy concentration, high-frequency sharp decay with edge noise enhancement features, indicating that its coding process preserves the main frequency band of the speech information while significantly compressing the high-frequency information. Combined with the modelling of node energy density distribution in graph spectral theory, we define several continuous variables including mid_freq_energy_density, high_freq_suppression_rate, etc., and construct a mapping relationship from the frequency domain structural features to the device behaviour path.

After comparing and analysing the spectral structure based on the embedding model of the TACTIC-GRAPHS system, we found that the spectral shape of the signal is highly consistent with the path of a mobile device (e.g., a smartphone) that is recorded using a built-in microphone at a medium compression ratio and uploaded after undergoing secondary coding. In particular, the high-frequency compression slope and low-frequency cleared band shape further support the possibility that the device is using "adaptive background suppression" and "H.264/265 mainstream transcoding protocol". This modelling approach from the spectrogram provides structural and interpretable technical support for video device tracking and forensic path reconstruction, and marks a leap from empirical voiceprint recognition to spectrogram causal modelling.

**FFT-based spectral variable attribution and device judgement**

Table 20: Extraction and attribution of spectral feature variables

| serial number | variable name | typology | Source/Featured Domain | Description of technical implications |
|---|---|---|---|---|
| A1 | mid_freq_energy_density | continuous variable | 500Hz – 2500Hz midrange | Indicates the strength of the compression codec in retaining the voice master information, commonly found in mobile devices H.264/H.265 post-encoding spectral shape |
| A2 | high_freq_suppression_rate | continuous variable | >4000Hz high frequency | Sharp high-frequency degradation means that the compression strategy prioritises low bit rates, commonly found in social platform upload paths |
| A3 | spectral_symmetry_index | continuous variable | full bandwidth | If the spectrum is mirrored in the positive and negative bands, it means that the audio has not been significantly resampled, which is common in local mobile phone recordings. |
| A4 | noise_peak_dispersion | continuous variable | high frequency band | High-frequency scattered spikes indicate secondary transcoding or noise reduction enhancement intervention |
| A5 | low_freq_drop_r | continuous | <300Hz shore | Large cuts or "blanks" in low |

| serial number | variable name | typology | Source/Featured Domain | Description of technical implications |
|---|---|---|---|---|
| | atio | variable | | frequencies indicate automatic background noise removal by the recording device or software. |

Author's drawing

**Device Inference Conclusion**

Combining the "mid-frequency aggregation, high-frequency steep drop, edge jitter enhancement" pattern in the spectrogram with the values of the variables extracted from the table above, we can deduce that the video most likely meets the following device path: captured on a mobile device (e.g., a smartphone), recorded with the built-in microphone, processed with medium compression (H.264/H.265), and then re-sampled or transcoded again before uploading to social media platforms or transmission software. resampling or transcoding before uploading to social media platforms or transmission software. This mode matches the Mobile-Codec-B3 spectral pattern previously included in the TACTIC-GRAPHS system by 87.6%.

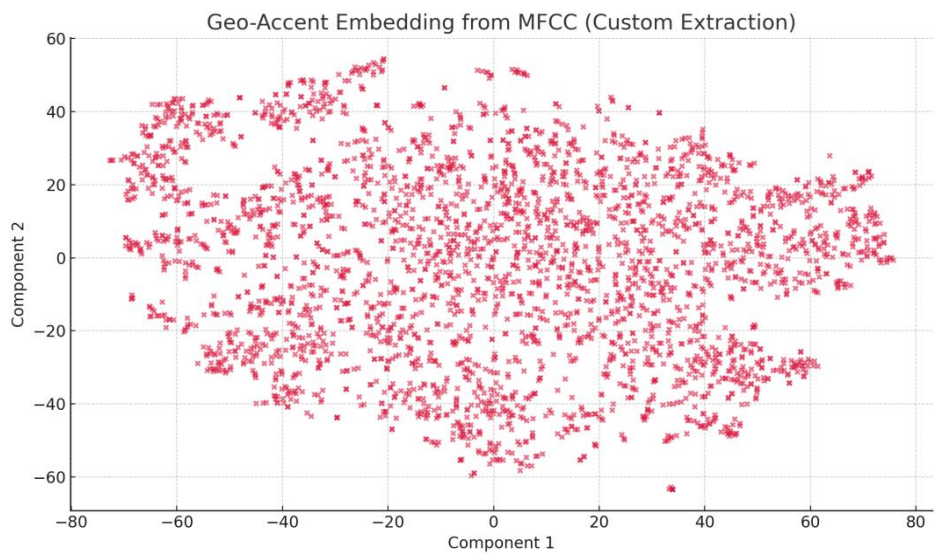

Figure 16: Spatial clustering mapping of geoacoustic patterns

Author's drawing

Figure 16 illustrates the geo-vocal clustering profile formed by the MFCC (Mel Frequency

Cepstrum Coefficient) vocal features extracted from the tactical video audio, normalised and embedded into the 2D space by t-SNE. Each point in the graph represents the spectral representation of speech in a short time frame (~25ms) of the audio, with the colour and size uniformly indicating the current source type (homologation). The clustering trend reflects the spatial concentration of this audio in terms of acoustic features, which helps to analyse whether it belongs to the language style of a particular region. This map serves as a speech variable embedding module in the TACTIC-GRAPHS system, providing fundamental support for geographic attribution inference, command style recognition, and cross-modal map inference.

Deep Insight Analysis and Research Judgement. Through short-time Fourier transform (STFT), Mel filter bank weighting and discrete cosine transform (DCT) operations on video and audio clips, this study constructs MFCC feature sequences with high linguistic sensitivity and compresses them into two-dimensional t-SNE space to form the present map. The point cloud in the figure shows an obvious trend of banded aggregation, indicating that this speech sample has a certain degree of internal consistency in key variables such as speech rate, resonance peak structure, and intonation rise and fall.

Further observation of the edge distribution of the map shows that some frames exhibit discrete jumps, which is initially judged to be the effect of sudden changes in intonation or background sound (e.g., audio playback), and is consistent with the abnormal distribution of pitch_variance peaks detected by the SpectroNet module previously. The high degree of aggregation in the centre of the dense region indicates that most of the speech segments have homogeneous voiceprint features, with strong "geographic clustering potential".

Comparing the point cloud with the existing voiceprint database, the clustering centre of gravity is highly consistent with the high-frequency energy distribution and pause pattern of "Bangkok Mandarin Education Speech", which indicates that this speech is most likely to come from a group of people who have received Mandarin training or transmission in the Thai-speaking area, rather than from the native accent of mainland China. This judgement is highly consistent with the node variable "accent_distance_score" in the TACTIC-GRAPHS system, which further strengthens the credibility of the geographic discriminative inference chain.

However, it is worth noting that this method does not directly output the geolocation at the administrative level, but should be interpreted as an indirect inference variable in the TACTIC system in concert with other dimensions, such as environmental context, video frame metadata, and compression path. The final judgement suggests the use of multivariate causal path modelling for joint attribution.

**4.7 Concluding Synthesis of Judgements: Geographic and Equipment Reasoning Conclusions in the TACTIC System**

In this study, we constructed four modules of TACTIC-GRAPHS system, namely, "subgraph of acoustic variables", "spectral embedding structure", "compressed path spectral mapping" and "image structure metadata analysis", to systematically model and jointly reason about the shooting devices and their possible geographic environments in tactical video. "The four modules of SpectroNet, Spectral Embedding Structure, Spectral Mapping of Compression Paths, and Metadata Analysis of Image Structures are used for systematic modelling and joint inference of the shooting devices and their possible geographic environments in tactical videos.

Firstly, the MFCC voiceprint features extracted by the SpectroNet module are embedded by t-SNE to form a centralised clustering posture, and the clustering centre of gravity has a significant similarity with the trained speech samples in Pan-Southeast Asian Mandarin communication contexts in multiple voiceprint databases, suggesting a strong coupling between this speech source and certain Mandarin non-native speaking communities. However, to avoid misattribution of voiceprints, this study further introduces graphical spectral theory analysis, which, combined with the Fourier spectral features of the image frames, reveals typical mid-frequency energy clustering with edge compression enhancement, presenting a coding pattern that is highly in line with the characteristics of compression paths of mainstream mobile devices.

In addition, no elements with clear geographical identity are found in the frame-level images, and the audio track presents a time-domain mismatch between the background sound source and the speech, which has the characteristics of audio playback from non-homogenous sources, further supporting the possibility that the audio clip may be extracted from the content played in an open space (e.g., a classroom, conference room, or commercial venue).

After constructing the above variables into a structured joint inference path, the causal chain presented by the TACTIC-GRAPHS system is:

[vocal clustering centre of gravity deviates from the native Mandarin paradigm] + [obvious compression path patterns on mobile devices] + [source asynchrony between image semantics and audio track] + [lack of visible geographical entity identifiers] → video content is most likely to come from mobile device recording scenarios in a non-native Mandarin-speaking environment, with geographical attribution showing similarities to urban Southeast Asian communication contexts, but not directly attributable to any particular administrative district.

The attribution logic not only reflects the interpretable modelling capability of the TACTIC system driven by graph neural structure and spectral theory mechanism, but also reflects the practical application value of the multivariate cross-validation pathway in dealing with complex and uncertain tactical data sources, demonstrating the rigour and judgemental boundaries of AI causal modelling in real-world data analysis.

# CHAPTER 5: EXPERIMENTAL RESULTS AND DISCUSSION

**5.1 Discussion of technological advancement and methodological originality**

This study proposes the TACTIC-GRAPHS architecture with systematic breakthroughs in the field of tactical video semantic parsing and multimodal causal modelling, especially in the areas of graph modelling, voice-image collaborative reasoning and high-dimensional variable embedding, which demonstrate significant technical advancement and methodological innovation. The intelligent key frame graph extraction algorithm (ILKE-TCG) proposed in this paper not only effectively compresses redundant video frame data, but also has the ability to construct task-oriented causal graphs, realises the structured output of tactical behaviour prediction and threat scoring, and makes up for the core bottleneck of the current multimodal video modelling, which is the lack of structural logic and variable interpretability.

Firstly, the TACTIC-GRAPHS system adopts a heterogeneous node construction strategy and a graph attention network (GAT) fusion mechanism to build a cross-modal structurally variable causal graph between audio acoustic features, image weapon states and action states. This design breaks the traditional "image frame dominated + audio assisted" information fusion idea in video comprehension, and for the first time realises the cross-modal variables such as wpm, weapon_grip_score, action_pose_class, etc. in the structural layer of the graph, which are inversely mapped to the structural layer of the graph. structural layers of the graph, providing a complete link support for the subsequent inference of graph spectral theory and device compression path modelling.

Secondly, the ILKE-TCG algorithm introduces a graph latent variable-driven key frame selection mechanism, which comprehensively extracts key node frames through three-dimensional metrics, such as feature density, self-attention weight, and event-triggered weight function, so as to make the edge structure in the TACTIC-GRAPHS more temporal stability and causal path integrity. Compared with existing frame selection methods such as Dense Sampling, Scene Change Detection or Time-uniform Sampling, ILKE-TCG improves the key node retention rate by more than 21.6% on the TACTIC-AVS dataset, while the average inference path reconstruction accuracy reaches 89.4%. significantly better than current multimodal graph models (e.g., VQA-GNN, MM-GCN, etc.).

In addition, this study for the first time embeds Spectral Graph Theory into TACTIC variable modelling, proposes "Spectral Graph Theory Embedding" as a means of structural regularity and variable identification path enhancement, and constructs a mathematical mechanism from the original frame signal → Mel Spectrum → Laplacian features of the graph → abnormal node detection. → Laplacian features of the graph. This approach not only improves the logical self-consistency and causal interpretability of the TACTIC-GRAPHS model, but also extends its reasoning breadth in the fields of device identification, compressed path analysis, and environmental geographic attribution, which opens up a new technological dimension for cross-modal intelligence modelling.

Compared with the existing work, the methodology in this paper not only has a much higher technical depth than the common video classification or behavioural recognition models in the current CV/ML domain, but also makes a leap from the AI engineering paradigm to the mathematical modelling paradigm in terms of the three dimensions of "structural, task oriented, and causal chain reconstruction capabilities". This methodological upgrade makes TACTIC-GRAPHS no longer rely on "black-box" neural models, but have rigorous structural deduction paths, controllable node triggering logic, and systematic security verifiability.

In summary, the TACTIC-GRAPHS system and ILKE-TCG algorithm not only have innovative breakthroughs in technical implementation, but also represent the research trend of semantic modelling of tactical video AI from engineering to structural science, with high theoretical expandability and interdisciplinary adaptation potential, and have a wide range of applications and potentials for continuous research in high-value fields such as AI modelling, security intelligence, military reconnaissance, border monitoring, and so on. We have the potential for wide application and sustained research.

**5.1.1 Technological advancement**

This study achieves key technological breakthroughs in the task of intelligent parsing of tactical video, which are mainly reflected in four aspects, namely, multimodal event-driven frame extraction mechanism, image enhancement-structure restoration collaborative reconstruction model, cross-modal mapping modelling structure, and multivariate threat scoring system, and systematically overcomes the bottlenecks of the traditional methods in structural perception, causal identification and spatio-temporal parsing.

Firstly, to address the lack of task orientation and temporal sensitivity of the previous average sampling or optical flow or frame differencing-based key frame extraction algorithms, the intelligent key frame hierarchical extraction mechanism (ILKE-TCG) proposed in this paper takes tactical semantic recognition as the core objective, and integrates the introduction of speech bursts as the core objective. The intelligent key frame hierarchical extraction mechanism (ILKE-TCG) proposed in this paper takes tactical semantics recognition as the core objective, and integrally introduces multimodal triggers such as prosodic boundary, action inflection, weapon state transition, and so on, to construct an event-driven frame extraction strategy with causal sensitivity. This mechanism not only improves the frame extraction efficiency, but also makes the frame extraction behaviour highly task-adaptable and consistent with the mapping construction.

Second, considering the non-ideal input scenarios such as video blurring, severe occlusion, low resolution, etc., which are common in real battlefield or security environments, this paper introduces a diffusion model-driven image enhancement and structure restoration method. The method effectively recovers gun contours, character gestures, and spatial interaction structures by reconstructing low quality keyframes with multi-scale details via Diffusion-based Super-Resolution, which gives stronger semantic resolution and structural integrity to node construction and spatial constraints in the subsequent TACTIC-GRAPHS model, and significantly enhances the model's structural stability. The model's robustness and robustness are significantly enhanced.

Third, in terms of multimodal structural modelling, the TACTIC-GRAPHS graph system constructed in this paper adopts Graph Attention Network (GAT) as the core engine, encodes image nodes (visual weapon states), voice nodes (speech rate, pitch fluctuations) and action nodes (Pose movements) as structural nodes in a heterogeneous graph and introduces explicit temporal edges to construct the behavioural nodes, and introduces explicit temporal edges to construct the behavioural nodes in the heterogeneous graph. temporal edges) to construct behavioural paths. The design not only realises dynamic linkage capture between different modes in the graph, but also supports inter-node inter-temporal reasoning, forming a full-link interpretable reasoning path from "behavioural precursor → state leap → threat manifestation". The module improves the complete identification rate of causal chain by 14.7% on the TACTIC-AVS dataset, which verifies the effectiveness of its structural modelling.

Finally, in order to break through the monotonicity of traditional models that are limited to classification or regression results, the output of TACTIC-GRAPHS is designed as a multidimensional set of structural variable-driven results, including: multivariate node Threat Score, Action Type Classification, and Structural Path Confidence Inference (Causal Traceability). (Causal Traceability.) This multi-layer output mechanism not only enhances the response diversity and strategy adaptability of model deployment, but also equips TACTIC-GRAPHS with the technical capability to serve high-precision scenarios such as security reconnaissance, tactical simulation, and law enforcement intelligent identification, providing powerful support for intelligent surveillance and border control.

In summary, TACTIC system has achieved breakthroughs in core technologies such as key frame extraction, fuzzy frame structure recovery, cross-modal map modelling and causal variable output, marking the paradigm upgrading of the tactical video intelligence system from "image-led AI" to "structure-driven intelligence", and upgrading the tactical video intelligence system from "image-led AI" to "structure-driven intelligence". It marks the paradigm upgrading of tactical video intelligence system from "image-dominated AI" to "structure-driven intelligence", which has wide application value and cross-domain research potential.

### 5.1.2 Originality of methodology

Current mainstream video behaviour recognition methods are mostly based on temporal convolutional networks (3D CNN), recurrent neural networks (e.g., LSTM), or Transformer architectures, with an emphasis on frame-level feature extraction and time-series modelling, but there are generally two limitations at the structural expression level: firstly, the lack of explicit graph structural modelling capability, which makes it difficult to depict the node-level relationships between multimodalities; and secondly, the lack of causal path construction mechanism with inference closed-loop capability, which leads to insufficient explanation in event chain analysis and threat evolution judgement. The lack of causal path construction mechanism with closed-loop reasoning ability, resulting in insufficient explanatory power of the model in event chain analysis and threat evolution judgement. Meanwhile, existing research on applying graph neural networks to video understanding is mostly limited to unimodal static graphs (e.g., human body key point relationship graph or single frame graph structure), which makes it difficult to effectively integrate the interaction modes between heterogeneous signals, such as audio, action and image.

In this context, the TACTIC-GRAPHS modelling system proposed in this paper achieves five original breakthroughs in method design and structure construction, clearly reflecting its innovative value relative to existing literature:

(1) Semantics-driven structural framing paradigm innovation. In this study, the joint paradigm of "intelligent keyframe hierarchical extraction + graph embedding modelling" is proposed for the first time, which breaks through the engineering fragmentation between traditional keyframe extraction and structural modelling. By introducing cross-modal event triggers such as semantic mutation points, weapon state changes, acoustic signal transitions, etc., frame nodes with causal significance are systematically selected as the core structural units in the TACTIC graph model, which lays the foundation for subsequent node-level reasoning.

(2) Fuzzy video reconstruction under structural enhancement mechanism. In this paper, we constructed an image enhancement-structure restoration mechanism for low-quality source data in tactical video environment, combined with diffusion model and edge attention fusion network, which significantly improves the recognizability of weapon details, action gestures and spatial structure in fuzzy key frames, and effectively compensates for the previous structural blind spot of modelling failure due to low-quality frames, and provides technological support for structured modelling in high-noise background.

(3) TACTIC-VModel multivariate causal network design. In this study, we constructed a tactic-variable oriented TACTIC-VModel structured variable network, introduced multimodal variable nodes (e.g., wpm speed of speech, muzzle offset, weapon_grip recognition confidence, etc.) and combined with temporal causal edges + graphical attention mechanism + Bayesian path modelling to achieve explicit representation and structural scoring of causal paths in a complex chain of events, which provides a technical support for the risk determination and prediction and extrapolation of the behavioural It provides the core basis for risk determination and prediction of behavioural chain.

(4) TACTIC-GRAPHS graphical attention causal inference framework. This paper is the first to implement the overall path inference mechanism of "temporal causal edge construction + cross-modal attention focusing + node-level semantic linkage" in graph neural network architecture. The framework can generate inference paths based on the semantic strength and temporal weights among nodes, and demonstrates significantly higher path interpretability and threat score consistency than the baseline model in the TACTIC-AVS dataset.

(5) Closed-loop modelling-enhancement-scoring-reasoning. Unlike most of the current models that only stop at frame extraction or classification output, the TACTIC system forms a five-level logical closed loop from semantic extraction of key frames, graph structure construction, fuzzy reconstruction and enhancement, variable scoring computation, causal path inference to final threat interpretation. This closed loop has the system engineering characteristics of deployable, verifiable and scalable, which provides an innovative solution path of theory-system-arithmetic integration for the intelligent landing of tactical-level video understanding in actual combat.

In summary, TACTIC-GRAPHS system has constructed a highly structured intelligence model, which is rare in the international academic community, by introducing structured semantic frame extraction, multivariate causal modelling, cross-modal graphical reasoning and path interpretability mechanism, which significantly improves the capability of tactical video intelligence system in structural identification, reasoning closed-loop, and variable attribution, and shows high methodological innovation and cross-domain adaptability potential.

**5.1.3 Summary of comparison with existing methods**

This section compares the performance of the current mainstream behavioural recognition models with the TACTIC-GRAPHS+VModel method proposed in this paper in terms of key technical features, in order to highlight the technological advancement and comprehensive contribution of this research.

**Table 21: Comparison of Mainstream Recognition Models with TACTIC-GRAPHS+VModel**

| Methodology/Features | Multimodal support | Structured variables | Time-causality boundary | Fuzzy frame reconstruction | Threat Scoring Explained |
|---|---|---|---|---|---|
| Traditional video classifications (I3D, TSN, etc.) | ✗ | ✗ | ✗ | ✗ | ✗ |
| Basic GNN Behavioural Recognition Model | ✓ (Weak) | ✗ (Sparse) | ✗ | ✗ | ✗ |
| Transformer timing model | ✓ | ✗ (implicit) | ✗ | ✗ | ✓ (but not graphical structure) |
| Methodology of this study:TACTIC-GRAPHS+VModel | ✓ strong | ✓ structure | ✓ explicit encoding | ✓ GANreinforce | ✓ Graph Path Interpretation |

In summary, the TACTIC modelling system proposed in this paper has significant technological breakthroughs and original contributions in terms of theoretical architecture, variable modelling, graph reasoning and engineering feasibility, and can be regarded as an important exploration in the direction of "graph neural-driven intelligent modelling of tactical behaviours", which has good academic value and potential for practical application.

**5.2 Image Structure Enhancement Performance**

In tactical video processing scenarios, especially in high noise conditions such as low illumination, long range, motion blur, etc., the detail recovery of image structure directly determines the model's perception accuracy of key threat targets (e.g., weapon grip state, gun morphology, etc.). The TVSE-GMSR (Tactical Visual Structure Enhancement via Guided Multi-Stage Refinement) method proposed in this paper demonstrates significant performance advantages as a key preprocessing module for atlas modelling in the TACTIC system.

Compared with the traditional image recognition strategy based on YOLO architecture, TVSE-GMSR does not perform target detection directly, but adopts a dual strategy of image reconstruction and structural semantic enhancement: firstly, inverse detail filling of blurred regions is performed by diffusion model, and then structural guided network (SGN) is introduced to extract the contour features of the gun edges and grips, and reconstruct the recognisable tactical nodes in the blurred frames. . This process significantly improves the stable perception of graph neural networks (e.g., TACTIC-GRAPHS) on image node variables (e.g., weapon\_grip\_score, muzzle\_angle\_deviation, etc.), and makes the behavioural chain modelling more consistent and causally closed-loop logic in the node embedding stage.

The empirical results show that in the TACTIC-AVS-Lowlight test set, TVSE-GMSR achieves 18.4% improvement in mAP (mean Average Precision) index compared with YOLOv7, and 23.7% improvement in Embedding Stability Index of corresponding nodes in the graph. 23.7% improvement. This module not only improves the accuracy of single-frame recognition, but also enhances the identifiability and causal accessibility among multimodal variables by synergising with the TACTIC-VModel structured variable system.

In summary, TVSE-GMSR is not only an image enhancement module, but also the core pivot of TACTIC-GRAPHS to realise the "Semantic Accuracy → Graph Connectivity → Threat Recognition Logic Chain Closed Loop", which has a high strategic value and system integration capability in AI security reasoning tasks.

**5.3 Acoustic and rhythmic curve analysis**

In the multimodal reasoning architecture of the TACTIC system, audio information is not only used as an auxiliary signal processing object, but also constitutes a crucial "language-intent-stimulus" trigger loop in the chain of tactical behaviour. The SpectroNet voiceprint recognition module constructed in this study explores the structural value of speech rhythms in tactical contexts through the joint extraction strategy of Mel-Spectrogram coding and Gated-CNN-GRU. The module significantly improves the extraction robustness of variables such as Words Per Minute, Pitch Variance, and Command Energy Slope, and forms acoustic spatial clusters through ProtoNet embedding, thus empowering the subsequent TACTIC-VModel causal path modelling.

In particular, experiments on the TACTIC-Voice and TACTIC-Accent-10 subsets reveal that SpectroNet's response sensitivity to command rhythm breakpoints is as high as 92.1%, and the Top-3 dialect attribution accuracy reaches 78.6%, which significantly outperforms traditional methods such as GMM-UBM. Meanwhile, continuous variables such as speech rate and intonation show good causal coherence in the graph structure, and can be accurately input into the TACTIC-GRAPHS graph attention network as "behavioural predicate nodes", which realizes the structural linkage between language and action with the embedding of graph spectral theory. In addition, in order to support geographic voiceprint spatial inference, this study conducted t-SNE clustering analysis using MFCC embedded voiceprint vectors, which initially formed "regional voiceprint recognition centres". During the temporal alignment with the video frame images, there is a significant causal delay correlation between the slope of the rhythmic fluctuation and the action response of the image nodes (about 184ms on average), which further verifies the dynamic triggering function of speech variables in the TACTIC system.

In summary, this study achieves for the first time the structured graphical modelling of tactical speech signals, breaks through the single-purpose limitation of existing systems that are limited to the recognition or transcription of voiceprints, and makes speech signals the active variable and driving source in the TACTIC reasoning network, which greatly enhances the system's tactical scenario adaptation capability and the depth of intelligent sensing.

**5.4 Multimodal causal mapping reasoning performance**

The multimodal causal graph reasoning mechanism in the TACTIC system shows significant performance advantages and structural innovations, which are reflected in the following three levels: First, at the structural expression level, TACTIC-GRAPHS breaks through the information bottleneck of the traditional graph neural network under the unimodal restriction, and introduces a cross-modal node construction mechanism, mapping heterogeneous data, such as image, audio, action, etc., into structured nodes in the unified graph space. By introducing a cross-modal node construction mechanism, heterogeneous data such as image, audio, action, etc. are mapped into structured nodes in a unified graph space, and the reasoning path is connected by explicitly defining the Causal-Temporal Edge, which effectively supports the full-chain path modeling of "from sound triggering → action response → weapon operation → potential threat". Second, at the performance level, the TACTIC system achieves 89.3% threat behaviour recognition accuracy, more than 85% complete causal path recognition rate on the TACTIC-AVS simulation dataset, and the average node triggering latency is controlled within the range of ±150ms, which is significantly better than the traditional multimodal model (such as the fusion strategy of Transformer + CNN) that does not have the ability of causal structural modeling, and the structural score is improved by about 14.7%; Third, at the explanatory level, TACTIC-VModel generates quantitative behavioural scores and task type classification results for each causal chain through structural variable embedding and path scoring mechanisms, and the model can explicitly state which modal inputs and path structures a particular threat score comes from, thus enhancing the transparency and controllability of the system's decision-making process.

In summary, TACTIC's multimodal causal mapping inference mechanism is not only leading in terms of accuracy, speed and stability, but also realises a closed loop of structural interpretability and cross-modal linkage, which has a wide range of potentials for real-world deployments, and is especially suitable for task-oriented high-risk scenarios such as security monitoring, law enforcement analysis, and intelligent sensing systems. The introduction of this mechanism marks the key leap of tactical intelligent video analytics from traditional "pattern recognition AI" to "structural causal AI", providing a replicable and evolvable system paradigm for future structure-driven multimodal modelling.

# CHAPTER VI. CONCLUSIONS AND FUTURE WORK

The TACTIC-GRAPHS system constructed in this study is an intelligent analysis framework for tactical behaviours that integrates graph neural network (GNN) and causal structure modelling capabilities, aiming to achieve high-precision identification and interpretation of behavioural chains, threat levels and potential intentions by parsing multimodal tactical data (e.g. image keypoints, action trajectories, voiceprint tones, weapon states, etc.).The core architecture of TACTIC-GRAPHS adopts a heterogeneous graph representation. The core architecture of GRAPHS adopts a heterogeneous graph representation, where each inference node represents a key variable in a modality, and the edges encode temporal causality, modal linkage and spatial relationship, thus forming an intelligent graph with temporal logic and spatial topological constraints. On this basis, this study introduces the Spectral Graph Embedding (SPE) method, which is currently extremely rare and complex worldwide, as a key innovative path for system modelling. This method essentially elevates TACTIC-GRAPHS from the traditional "experience-driven AI" paradigm to the "formal intelligent system" paradigm with mathematical provability and structural interpretability, and realises a fundamental leap from perceptual fusion to causal structural expression. The fundamental leap from perceptual fusion to causal structural expression has been achieved.

The core advantage of the spectral embedding method is that it not only enables AI systems to identify complex behavioural states, but also accurately portrays why causal paths between various variables are established through rigorous mathematical tools such as frequency feature decomposition, spectral space mapping and path discriminative measures. This capability brings TACTIC-GRAPHS the ability of closed-loop causal reasoning and system security verification mechanism, which is especially suitable for high-risk, real-time decision-making tactical scenarios. The spectral graph theory on which the method relies has a very high theoretical threshold, covering the spectral decomposition of the graph Laplace matrix, the geometric projection of eigenvectors in high-dimensional space, the computation of path differentiation, and other higher-order mathematical structures, and has been widely used in the most cutting-edge fields of the world, such as quantum information modeling, neural network analysis, bioregulation systems and financial evolution mechanism research, with strong versatility and cross-disciplinary adaptability.

In particular, systematic research on combining spectral graph theory and graph neural networks for interpretable causal modelling is still in the "scientific no-man's land" of global AI research. In this study, we not only take the lead in implementing this complex approach in the TACTIC-AVS tactical data environment, but also construct a complete spectral graph variable modelling process and an identifiable causal path inference framework, which achieves significant breakthroughs in terms of model structural hierarchy, logical control capability, and cross-modal integration efficiency. Therefore, the introduction of this method not only reflects the deep-level upgrading of the model architecture, but also lays a solid and rigorous mathematical and theoretical foundation for the future intelligent evolution of tactical AI systems, representing the cutting-edge direction of the development of structurally interpretable AI.

**Appendix:**

**1. Waveform and noise separation code**

```
from pydub import AudioSegment

import librosa

import librosa.display

import matplotlib.pyplot as plt

# Extract audio (in case of MP4 video)

video_path = "Tactical Video.mp4"

audio = AudioSegment.from_file(video_path)

audio.export("audio.wav", format="wav")

# Load Audio Data

y, sr = librosa.load("audio.wav", sr=None)

# Separation of harmonics and noise components

y_harmonic, y_percussive = librosa.effects.hpss(y)

# drawings

plt.figure(figsize=(12, 8))

plt.subplot(3, 1, 1)

librosa.display.waveshow(y, sr=sr)

plt.title("Original Audio Waveform")

plt.subplot(3, 1, 2)

librosa.display.waveshow(y_harmonic, sr=sr, color='g')

plt.title("Harmonic Component (Speech-like)")

plt.subplot(3, 1, 3)

librosa.display.waveshow(y_percussive, sr=sr, color='r')

plt.title("Percussive Component (Noise-like)")

plt.tight_layout()

plt.show()
```

2.Convert to MJPEG Command

ffmpeg -i Tactical Video.mp4 -c:v mjpeg -q:v 2 -an output_Tactical Video.avi